%% file: main.tex
\documentclass[lettersize,journal]{IEEEtran}

\usepackage{algorithmic}
\usepackage{algorithm}
\usepackage{array}
\usepackage[caption=false,font=normalsize,labelfont=sf,textfont=sf]{subfig}
\usepackage{textcomp}
\usepackage{stfloats}
\usepackage{url}
\usepackage{verbatim}
\usepackage{graphicx}
\usepackage{cite}

\usepackage{siunitx}

\usepackage{amsmath,amssymb,amsfonts,fixmath}

\usepackage{graphicx}
\usepackage{tabularx,booktabs}
\usepackage{diagbox}
\newcolumntype{C}{>{\centering\arraybackslash}X}
\usepackage{makecell}
\newcommand{\Mod}[1]{\ (\mathrm{mod}\ #1)}

\begin{document}

\title{IRS-enabled Breath Tracking with Colocated Commodity WiFi Transceivers\\
	\thanks{This work was funded by the Federal Ministry of Education and Research (BMBF) of the Federal Republic of Germany (F\"orderkennzeichen 16KIS1152, mINDFUL and 16KIS1235, MetaSEC).}
}

\author{Simon~Tewes$^{\ast }$, Markus~Heinrichs$^{\dagger}$, Rainer~Kronberger$^{\dagger}$, Aydin~Sezgin$^{\ast }$\\
	$^{\ast }$Chair of Digital Communication Systems, Ruhr-University Bochum, Bochum, Germany,\\
	$^{\dagger}$High Frequency Laboratory, TH Cologne - University of Applied Sciences, Cologne, Germany\\
\{simon.tewes, aydin.sezgin\}@rub.de,\{markus.heinrichs, rainer.kronberger\}@th-koeln.de}

% The paper headers
\markboth{Submitted to IEEE Internet of Things Journal}%
{Shell \MakeLowercase{\textit{et al.}}: IRS-enabled Breath Tracking with Colocated Commodity WiFi Transceivers}

\maketitle

\begin{abstract}
Intelligent reflecting surfaces (IRS) are a key enabler of various new applications in 6G smart radio environments. By utilizing an IRS prototype system, this paper aims to enhance self-interference (SI) cancellation for breath tracking with commodity WiFi devices. SI suppression is a crucial requirement for breath tracking with a single antenna site as the SI severely limits the radio sensing range by shadowing the received signal with its own transmit signal. To this end, we propose to assist SI cancellation by exploiting an IRS to form a suitable cancellation signal in the analog domain.
Building upon a 256-element IRS prototype, we present results of breath tracking with IRS-assisted SI cancellation from a practical testbed. We use inexpensive WiFi hardware to extract the Channel State Information (CSI) in the 5~GHz band and analyze the phase change between a colocated transmitter and receiver with added local oscillator (LO) synchronization. We are able to track the breath of a test person regardless of the position in an indoor environment in a room-level range.
The presented case study shows promising performance in both capturing the breath frequency as well as the breathing patterns.
\end{abstract}


\section{Introduction}
\IEEEPARstart{W}{ith} 
the demographic change in society, more and more people are dependent on care and have certain health conditions \cite{PopulationReport}. In order to enable the affected persons to live independently in their own homes for longer, a wide variety of systems for monitoring vital parameters have been developed \cite{AALsurvey}. This kind of support is called Ambient Assisted Living (AAL) and can be used for monitoring vital parameters like respiration as well as for fall detection and many more. Respiration and heartbeat are useful vital signs for physical health monitoring as the information provides important indications of medical problems \cite{SidSyndrom}. In this paper, we focus on respiration detection as a variant of AAL.

Traditional methods of respiration detection, often use body-worn sensor technology that detects the expansion of the chest \cite{AALsurvey}. These sensor technologies are intrusive and often perceived as uncomfortable, so they are not suitable for long-term respiration monitoring \cite{BreathTrack}.

Due to advances in health monitoring research related to the Internet of Things, some previous works have already presented contactless variants of respiration detection using radio waves \cite{BreathTrack, PhaseBeat, ContactlessRespirationPeng, MonitoringVitalSignsYang, WangIoT}.  
The previous works either use specialized hardware with directional antennas \cite{SmartHomesKatabi} or exploit a separation of multiple nodes in the experiment space, similar to bistatic radar, with the test subject in between transmitter and receiver \cite{BreathTrack, PhaseBeat, ContactlessRespirationPeng, MonitoringVitalSignsYang, WangIoT,SmartphoneRespIoT}. However, this special placement of devices and test subjects is difficult to implement in practice, and also concepts that require a cabled connection of two nodes as a reference limit the use of this technique, since coax cables have to be laid from one end of the room to the other.

Among many other promising applications and new techniques, research on 6G investigates the use of IRS. The IRS is a synthetic surface able to passively manipulate the reflected signal depending on an electrical reconfiguration of the elements. For the first time, this promising concept allows to deviate from the paradigm of a determined, passive communication channel and rather allows to optimize the channel for a specific application or user. These innovative features qualify the IRS to be a part of an upcoming communication standard \cite{IRSsurvey}. In view of this development, IRS will become more readily available and part of the communication infrastructure in the future. 

Therefore, in this paper, we propose an Intelligent reflecting surface (IRS) for simplifying the user equipment (UE) hardware in full-duplex (FD) communications as well as joint communication and sensing contexts. 
In in-band full-duplex communication, a frequency ressource is used for simultaneous transmission and reception on the same carrier frequency. This simultaneous use of a frequency resource increases spectral efficiency by up to a factor of two, i.e. the throughput of a communication link is doubled compared to classic half-duplex systems.
The application of joint communication and sensing is closely related to full-duplex communication. In this field, a communication system is used for simultaneous transmitting data and sensing. During data transmission, the reflection of the transmitted signal from the environment or the target is received simultaneously. Thus, additional sensors can be omitted and an added value can be added on top of the communication. 

In such systems, the main challenge is to suppress the self-interference (SI) caused by simultaneous transmission and reception on one frequency. This is classically achieved by RF hardware such as circulators or auxilary transmitters. Addressing these issues, we propose a novel concept which utilizes an IRS to form a cancellation signal that is phase inverted to the sum of the individual self-interference components caused by multipath reflections and hardware imperfections such as limited antenna isolation and internal leakage in a FD transceiver. The signal reflected from the IRS aims to cancel the SI at the receive antenna of a node. By using a greedy algorithm to configure the IRS, our setup adapts to various indoor multipath environments and is able to generate the desired cancellation signal without prior channel or location knowledge. Furthermore, by taking advantage of the IRS infrastructure that will be deployed in 6G networks, the implementation costs for a single UE is reduced. The reduced SI achieved in this way allows us to simultaneously transmit a WiFi packet on a wifi channel from one antenna location, as well as receive it back at the same location.

Since the IRS-enabled SIC cancels static reflections, leakage and other quasi-static signal components, the total signal power at the receiver stemming from the SI is reduced. By reducing the SI power, the signal of interest becomes is lifted out of th quantization noise. Thus, a much more detailed channel estimation is achieved. If we now consider an indoor respiration detection scenario, the raising and lowering of a subject’s chest causes a time-varying small phase change on some multipath paths. Initially, these small changes were overshadowed by the SI and disappeared in the quantization noise of the channel estimation. With the quasi-static SI components now removed, the AGC is able to better amplify the received signal into the ADC range, reducing the quantization noise and the small phase variations caused by reflection from a subject’s chest can be detected from only one antenna location, largely independent of the subject’s position in space.

\subsection{Prior work}
In the past several concepts for breath tracking with commodity WiFi hardware have been shown. In this section, we briefly introduce the core concepts of breath tracking and current state-of-the-art. In table~\ref{tab:RelatedWork} we give an overview of prior work in comparison to our proposed approach in important key aspects, which we further elaborate below.

	\begin{table*}[!ht]
		\caption{Related Work Comparison}
		\label{tab:RelatedWork}
		\begin{tabularx}{\linewidth}{l|*{15}{C}c@{}}
			\toprule
			\diagbox[width=3.5cm, height=1.5cm]{\raisebox{3pt}{\hspace*{0.1cm}Key aspect}}{\raisebox{-0.7cm}{\rotatebox{90}{Paper}}}  & \rotatebox[origin=c]{90}{Our approach} & \rotatebox[origin=c]{90}{BreathTrack \cite{BreathTrack}}  & \rotatebox[origin=c]{90}{PhaseBeat \cite{PhaseBeat}}  & \rotatebox[origin=c]{90}{Liu, X. et al.\cite{ContactlessRespirationPeng}} & \rotatebox[origin=c]{90}{Liu J. et al.\cite{MonitoringVitalSignsYang}}  & \rotatebox[origin=c]{90}{Wang et al.\cite{WangIoT}} & \rotatebox[origin=c]{90}{Adib et al. \cite{SmartHomesKatabi}} & \rotatebox[origin=c]{90}{MoBreath \cite{SmartphoneRespIoT}}  & \rotatebox[origin=c]{90}{TensorBeat \cite{TensorBeat}} & \rotatebox[origin=c]{90}{FullBreathe \cite{FullBreathe}} \\ \midrule
			Number of TX/RX antennas & \makecell{1 TX \\ 1 RX} & \makecell{1 TX \\ 2 RX}  &\makecell{1 TX \\ 2 RX} & \makecell{1 TX \\ 3 RX} & \makecell{1 TX \\ 1 RX} &\makecell{3 TX \\ 3 RX} &\makecell{1 TX \\ 1 RX} &\makecell{ 1 TX \\ 1 RX} &\makecell{1 TX \\ 3 RX} &\makecell{1 TX \\ 2 RX} \\ 
			Omnidirectional antennas  &\checkmark & \checkmark & \checkmark & \checkmark & (x) & \checkmark & \checkmark & \checkmark & x & \checkmark\\ 
			RF reference channel & x & \checkmark & x & x & \checkmark & x & x & x & x & x\\ 
			% \addlinespace
			Transceiver hardware & WiFi & WiFi  & WiFi & WiFi & WiFi  & WiFi & FMCW & WiFi & WiFi & WiFi \\ 
			Synchronization & LO & RF  & x & x & x & x & inherent & x & x & x\\ 
			Colocated transceivers & \checkmark & x  & x  & x & x  & x& \checkmark  & x & x & x \\ 
			Free position of test person  & \checkmark  & x  & x  & x  & x & x & x & (\checkmark) & x & \checkmark \\ 
			\bottomrule
		\end{tabularx}
	\end{table*}

In \cite{BreathTrack}, Zhang et al. demonstrated respiration detection with commodity WiFi hardware in an indoor scenario using a wired reference channel. For this purpose, the transmit signal was split by an RF splitter and passed once to a transmit antenna and once to the receiver through a coaxial cable. The receiver has three receive chains, one of which receives the reference signal and the other two are connected to two antennas. The reference signal enables a correction of the carrier frequency offset (CFO) and a measurement of the phase information over several WiFi packets. Although this setup shows good performance in the presented scenarios, it requires an additional wired reference channel in the carrier frequency range and at least 3 receiver chains, as well as an RF splitter. Furthermore, the setup is operated with spatial separation of the transmitter and receiver, so that the reflection from the chest is large enough relative to the direct path. 
Our approach, on the other hand, uses colocated antennas and operates with a single transmit and a receive antenna and associated transceiver chain, hence reducing the cost and complexity of the user equipment (UE). Another advantage of our approach is the simpler deployability, by utilizing only one antenna location, which eliminates the use of expensive coaxial cables in the carrier frequency range.

In \cite{PhaseBeat}, Wang et al. have shown a scheme where two WiFi devices are spatially separated, with the test subject located between the transmitter and receiver. The phase difference at two receiving antennas is used to obtain stable phase information containing the breath information. This setup, however, requires at least two receive antennas and is also operated only with spatial separation between transmitter and receiver. Additionally, this setup puts preliminaries on the position of the test subject, resulting in limited applicability.
Our approach allows the use of a single antenna site and thus removes the preliminaries to the test subject's position to the greatest extent possible, resulting in a much more convenient application. In addition, the reduction to one antenna location and the use of a single transmitting and one receiving antenna reduces the cost of a UE.

In \cite{SmartHomesKatabi}, Adib et al. present a FMCW radar approach on breath tracking. The approach uses a proprietary radar that sweeps from \SI{5.46}{\giga\hertz} to \SI{7.25}{\giga\hertz}, by means of two directional antennas aimed at the subject's chest, the distance to the subject, or the elevation and depression of the chest, can thus be measured. This approach, although providing good measurement results, is rather impractical, since the use of directional antennas determines the position of the test subject. In addition, it requires the use of specialized hardware that is only responsible for sensing, thus driving up the cost of a device.
In our approach, we utilize commodity WiFi, adding value through reusability as communication hardware or in the joint communication and sensing context. In addition, we use omnidirectional antennas that do not restrict the position of the test subject.

All of these previously outlined concepts use either specialized RF hardware \cite{BreathTrack, SmartHomesKatabi}, directional antennas \cite{SmartHomesKatabi, MonitoringVitalSignsYang}, or require spatially separated transmitter and receiver nodes \cite{BreathTrack,PhaseBeat,MonitoringVitalSignsYang,ContactlessRespirationPeng, FullBreathe}. This is why we propose to utilize IRSs in future 6G networks to assist the SIC and relax the requirements to specialized hardware on the UE side. This can significantly reduce costs and save complexity at the UE. By using channel estimation, the IRS is tuned so that the reflected signal cancels the SI at the UE and thus enables breath tracking from a single antenna site.

The remainder of the paper is structured as follows. First, in Section~\ref{sec:ChannelModel} we introduce the channel model. In Section~\ref{sec:IRSsupression} we propose our approach to IRS assisted SI cancellation. Next, we detail a greedy algorithm to optimize the IRS state to minimize the SI. In Section~\ref{sec:CSI} we elaborate on the CSI data used in WiFi transceivers, which is used for data aquisition in our experiments.
Following, in Section~\ref{sec:ExperimentStudiesSetup} we first describe the experiment layout and the scenario setting. Then, we detail the used IRS prototype and the transceiver hardware. Finally, in Section~\ref{sec:Results} the results of our experiments are shown.

\input{ChannelModel.tex}

\input{IRSsiSupression.tex}
\input{GreedyAlgorithm.tex}

\input{wifiCSI.tex}

\section{Experimental Studies setup}\label{sec:ExperimentStudiesSetup}
In this section, we first describe the measurement setup for the following breath tracking experiments. Then, we describe the individual hardware components of the setup in detail and explain the parameters used for our investigations.

\input{experimentFloorSetup.tex}
\input{transceiver.tex}

\input{irs.tex}

\section{Results}\label{sec:Results}
In this section, we present the results of our investigations. First, we show the phase stability of the applied measurement system with the added LO synchronization over an extended ambient temperature range. Then we show the performance of the presented greedy algorithm and finally we present the results in the application of breath tracking.

\input{phaseStability.tex}

\input{algoConvergence.tex}
\input{breathTracking.tex}

\section{Conclusion}
In this paper, we have demonstrated for the first time that IRS-assisted SI cancellation with commodity WiFi devices for respiration detection is possible with a single colocated antenna site.
We have built a prototype system that has delivered very promising results in the \SI{5}{\giga\hertz} WiFi band in a typical indoor environment. We first investigated the reduction of SI using a greedy algorithm to tune the IRS, and ensured the phase stability of the measurement system using LO synchronization. This made it possible with only a single antenna site, with two antennas separated by $\lambda/2$ from each other, to track a phase change in the CSI data over a period of several minutes. It was possible to detect not only the breathing rate of a test subject, but also the pattern of breathing.
The presented approach has shown a very good performance, which can significantly increase the quality of life in ambient assisted living scenarios, since body-worn devices for monitoring vital parameters can be omitted.

\bibliographystyle{IEEEtran}
\bibliography{./bib} 

\end{document}

%% file: ChannelModel.tex
\begin{figure}[!t]
	\centering
	\includegraphics[width=\linewidth]{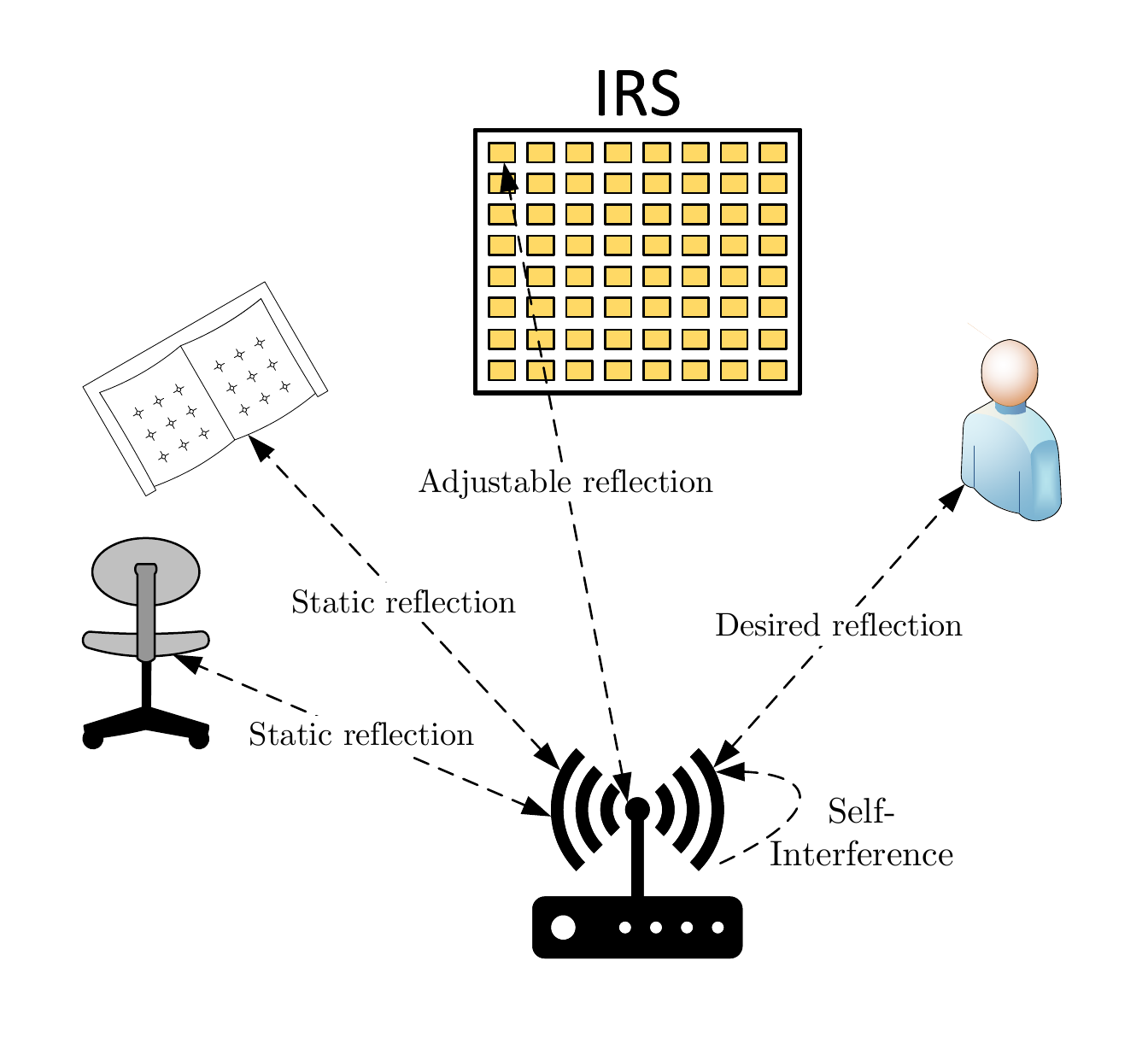}
	\caption{Scenario overview}
	\label{fig:ScenarioSetting}
\end{figure}

\section{Channel Model} \label{sec:ChannelModel}
The acquisition of vital parameters of a test subject rarely takes place under idealized conditions where only the direct path exists. It is therefore necessary to establish a suitable channel model for a complete representation of the conditions. In indoor multipath environments, a variety of propagation paths exist that are affected by the environment in distinct ways. In Fig.~\ref{fig:ScenarioSetting} the propagation paths in a typical indoor scenario are illustrated. The present paths can be divided into four types, which are explained in detail in the following and finally summarized into an overall channel model.

First, we consider the reflection of interest that contains the relevant change due to breathing. The detection of a test subject's breathing by means of radio signals uses the premise that breathing is always associated with a periodic raising and lowering of the test subject's chest. When a radio wave hits the subject's chest, it is reflected and can be evaluated with a suitable receiver. 
The path length $d(t)$ that a radio wave travels from a transmitting antenna across the chest to the receiving antenna changes as a function of the deflection of the chest.
If only the direct line-of-sight path is considered with a test subject sitting quietly, the influence of breathing on the path can be given by a complex attenuation $a_0$ and a phase term, as the channel state information $y_d(t)$ as follows
\begin{align}
	y_d(t)=a_0\cdot e^{-j2\pi\frac{d_0+d(t)}{\lambda}}.
\end{align}
Here $d_0$ represents the time-invariant part of the path length, $d(t)$ the part of the path length variable by the elevation and depression of the thorax, and $\lambda$ the wavelength of the carrier frequency used.
It can be seen that the change in chest expansion is expressed in a phase change of the channel state information $y_d(t)$ of the direct path and depends on the expansion of the chest, as well as the carrier wavelength. The typical elevation of the thorax $max(|d(t)|)$ for healthy adults is on average about \SI{30}{\milli\meter} for deep breathing and \SI{3}{\milli\meter} for shallow breathing, whereas Kaneko et al. \cite{BreathExpansionStudy} demonstrated a dependence on age, sex, and weight of the test subject's. Thus, for a realistic scenario using 5 GHz carrier frequency, a maximum phase change of the direct path of {$\Delta\phi=\frac{2\cdot max(|d(t)|)}{\lambda}=\frac{2\cdot\SI{30}{\milli\meter}}{\SI{60}{\milli\meter}}=360^{\circ}$} for deep breathing and 36° for shallow breathing is expected.

In addition to the direct paths altered by respiration, there is also a large proportion of static paths in multipath environments. These static paths result from the reflection of the transmitted signal from boundary surfaces, such as ceilings, walls, floors, furniture, and interior in general. These objects are stationary for a typical period of several seconds to hours and are not moved. Thus, the static paths can be assumed to be constant during the acquisition period under consideration. However, because boundary surfaces such as walls are very large, these paths make up a very large fraction in terms of received power. Static paths can be represented by a complex attenuation $a_l$, as well as a phase term depending on the path length $d_l$ and the carrier wavelength $\lambda$. The effect of all static reflections can be represented as a sum of all paths as follows
\begin{align}
	y_s=\sum_{l=1}^{L}a_l\cdot e^{-j2\pi\frac{d_l}{\lambda}}.
\end{align}
Hereby, $L$ denotes the total number of static paths, the number of which is largely determined by the number of scatterers in the near environment.  

In addition to the passive objects and their contributions to the reflections discussed above, in future 6G networks, adjustable radio environments will be available, allowing a radio channel to be reprogrammed by the users or network operator's equipment. A promising candidate for adjustable radio environments is the IRS with binary switching states. These surfaces, which are divided into patches, allow the phase of a signal reflected from that patch to be changed. These adjustable reflection paths must be accounted for in the channel model. The reflection of a signal at the $i$-th patch of the surface can be formulated as a function of the binary switching state $s_i$ by a complex attenuation factor $a_i$, a phase term and the switchable phase component of the surface. A summation over all $I$ elements of the surface thus gives the CSI for the IRS-assisted paths as
\begin{align}\label{equ:yIRS}
	y_\text{IRS}(s)=\sum_{i=1}^{I}a_i\cdot e^{-j2\pi\frac{d_i}{\lambda}}\cdot e^{js_i\phi}.
\end{align}
Here $\phi$ is the phase shift capability of the IRS, a parameter determined by the design of the surface. Usually it is 180° to achieve a maximum effect on the reflected signal.

In addition to the paths affected by the propagation environment described above, there are always paths from the transmitter to the receiver that are independent of the propagation environment. This direct crosstalk of the transmit signal into the receive path is generally referred to as self-interference and is caused by, among other things, internal leakage in the transceiver, limited analog isolation of transmit and receive antennas or, in the case of a common transmit/receive antenna, limited isolation of circulators, and antenna mismatch. This self-interference occurs in any full-duplex node and therefore must be specifically addressed. This self-interference is usually the strongest path in terms of power in colocated nodes. The path is composed of a variety of parameters, which can usually be assumed to be time-invariant over the time period under consideration.\cite{DuarteFD,LiuFD} The SI is additionally temperature dependent, but this is also negligible for many applications due to the slow temperature change of components. The self-interference can be represented accordingly as follows
\begin{align}
	y_\text{SI}=a_\text{iso}\cdot e^{-j2\pi\theta}.
\end{align}
Here $a_\text{iso}$ denotes the complex attenuation, which results from the antenna isolation, as well as internal leakage of the transceiver and cable attenuation. The phase term $e^{-j2\pi\theta}$ describes the associated phase, which strongly depends on the cable lengths, antenna delays and the geometrical arrangement of the antennas. For the narrowband case these parameters can be summarized with the described representation, since the determination of the individual components of the SI is often not possible and also not necessary.

The model for the entire channel can thus be set up as a sum of the individual paths as 
\begin{align}\label{equ:ChannelModel}
	y(S,t)=y_s+y_d(t)+y_\text{IRS}(S)+y_\text{SI}.
\end{align} 
Higher-order reflection paths with more than one reflection are not accounted for in the channel model, as they are negligible due to the very low received power from it \cite{HighOrderMultipath}.

%% file: IRSsiSupression.tex
\section{IRS-assisted SI supression}\label{sec:IRSsupression}
From the presented channel model it is apparent that the change in $y_d(t)$ caused by the subject's breathing accounts for only a small part of the received signal. The received power components due to static reflections and self-interference are orders of magnitude larger than the single reflection of the subject's chest. As a result, overshadowing of the phase change caused by respiration by the other signal components occurs. Considering finite ADC resolutions of communication systems, this change thus falls into the range of quantization noise and can no longer be recovered \cite{FDquantizationNoise}. 
Although transceiver and antenna design can already achieve a good amount of SI suppression, this is still insufficient for detecting extremely small signals such as respiration. In particular, when using commodity WiFi or other low-cost transceivers, the ADC resolution of the systems is limited and extremely small changes in the overall signal cannot be detected. Therefore, in previous WiFi CSI based work \cite{BreathTrack, PhaseBeat, ContactlessRespirationPeng, MonitoringVitalSignsYang}, two spatially separated transceivers have always been used to circumvent the problem of self-interference and to magnify the effect of breathing in the received signal by organizational measures.
Since this approach is hardly applicable in practice, we propose to use the IRSs available in future 6G networks to further reduce the effect of the SI and achieve single-node coverage.
From \eqref{equ:ChannelModel} it can be seen that the static reflections and the self-interference are time-invariant and not adjustable. The IRS reflection, on the other hand, is adjustable by the IRS state matrix $\mathbold{S}$. The property of IRS to change the phase of reflection paths can be used to manipulate a reflected signal to be phase-inverted to the sum of the self-interference and static reflections.
Since it is known that the self-interference and static reflections are time-invariant over the measurement period, cancellation of the static components can be achieved by selecting an appropriate state matrix $\mathbold{S}$ of the surface, while leaving the time-varying component of the respiratory signal unaffected. The cancellation of the static components result in a much larger phase change, which is caused by the respiration and is thus raised out of the quantization noise.
If we consider the sum-receive power components, in the form of a link-budget calculation, assuming that by a suitable choice of the IRS state $\mathbold{S}$, $y_\text{SI}$ is exactly phase inverted to self-interference and perfect cancellation results, then it follows for the residual SI power $P_\text{res}$
\begin{align}\label{equ:power}
	P_\text{res}=P_\text{tx}-\alpha_\text{iso}+P_{s}-P_\text{IRS},
\end{align}
where $P_\text{tx}$ is the transmit power, $\alpha_\text{iso}$ ist the antenna isolation given by the antenna assembly, $P_s$ is the power received from static reflections and $P_\text{IRS}$ is the power reflected from the IRS to the receive antenna.  (all values, except $\alpha_\text{iso}$ in dBm).
From \eqref{equ:power} it is visible that the received self-interference power depends only on noninfluenceable powers except for $P_\text{IRS}$. Thus, for maximum self-interference cancellation, the 180° phase-inverted signal reflected from the IRS must have a power exactly equal to the sum of the powers of $P_\text{tx}-\alpha_\text{iso}$ and $P_s$ for them to exactly cancel. Too high power of the IRS reflection would, despite correct phasing, have a negative effect, because it would overshoot the target.
The goal now is to find a suitable state of the surface for which the condition of 180° phasing, relative to the SI is given. We therefore consider the $2\pi$-periodic phase on the IRS-affected reflection paths. Since a beamforming effect of IRS occurs only in the far-field \cite{IRSnearfieldFarfieldBjoernson}, near-field or transition field conditions can be assumed in indoor environments in view of large dimensional IRS. With the array far-field radius given by \cite{IRSfarField} as
\begin{align}
	r_{FF}=\frac{2D^2}{\lambda},
\end{align} 
where $D$ is the longest dimension of the IRS, and $\lambda$ is the carrier wavelength, the far-field boundary $r_{FF}$ of the entire IRS is calculated. For a small surface with an edge length of \SI{40}{\centi\meter} and a carrier frequency of \SI{5}{\giga\hertz} the far-field begins at a distance of \SI{5.33}{\meter}. Since the far-field boundary increases quadratically with the edge length of the surface, a near-field or transition field can always be assumed for large-dimensional IRS in indoor scenarios. We therefore use the element-wise representation of the IRS also given in \eqref{equ:yIRS}, because the antenna array distance in indoor scenarios is always larger than the far-field distance for an individual element of the surface.

In sum of all phase changes introduced by the IRS we get the sought cancellation signal. If the distances and geometrical dimensions of antennas and surfaces relative to each other are known, the phase of the respective path can be calculated by a simple geometrical calculation of the path length to the respective patch of the surface as
\begin{align}
	d1=\sqrt{d^2+\Bigl(d_h\cdot\frac{w}{N_c}\Bigr)^2}\\
	d2=\sqrt{d_1^2+\Bigl(d_v\cdot\frac{w}{N_r}\Bigr)^2}\\
	\phi_i=\frac{2\pi}{\lambda}\cdot\bigl(d_2\Mod{\lambda}\bigr).
\end{align}
 Here, $d$ is the distance between transmit/receive antennas and the surface, $w$ is the width of the surface, and $h$ is the height of the surface. By applying the geometric relations with the known position of the antenna and a relative indexing of the horizontal and vertical patches $d_h$ and $d_v$ respectively, the absolute path length can be calculated. Thus, the phase of the $i$-th patch is obtained as $\phi_i$ by a modulo division by the wavelength $\lambda$. The antenna spacing for separate transmit and receive antennas is usually negligible for the calculation of the phase distribution, since the distance of the antennas to each other $d_a$ is much smaller than the distance of the antennas to the surface $d_\text{IRS}$.
\begin{figure}[!t]
	%\centering 
	\includegraphics[width=\linewidth]{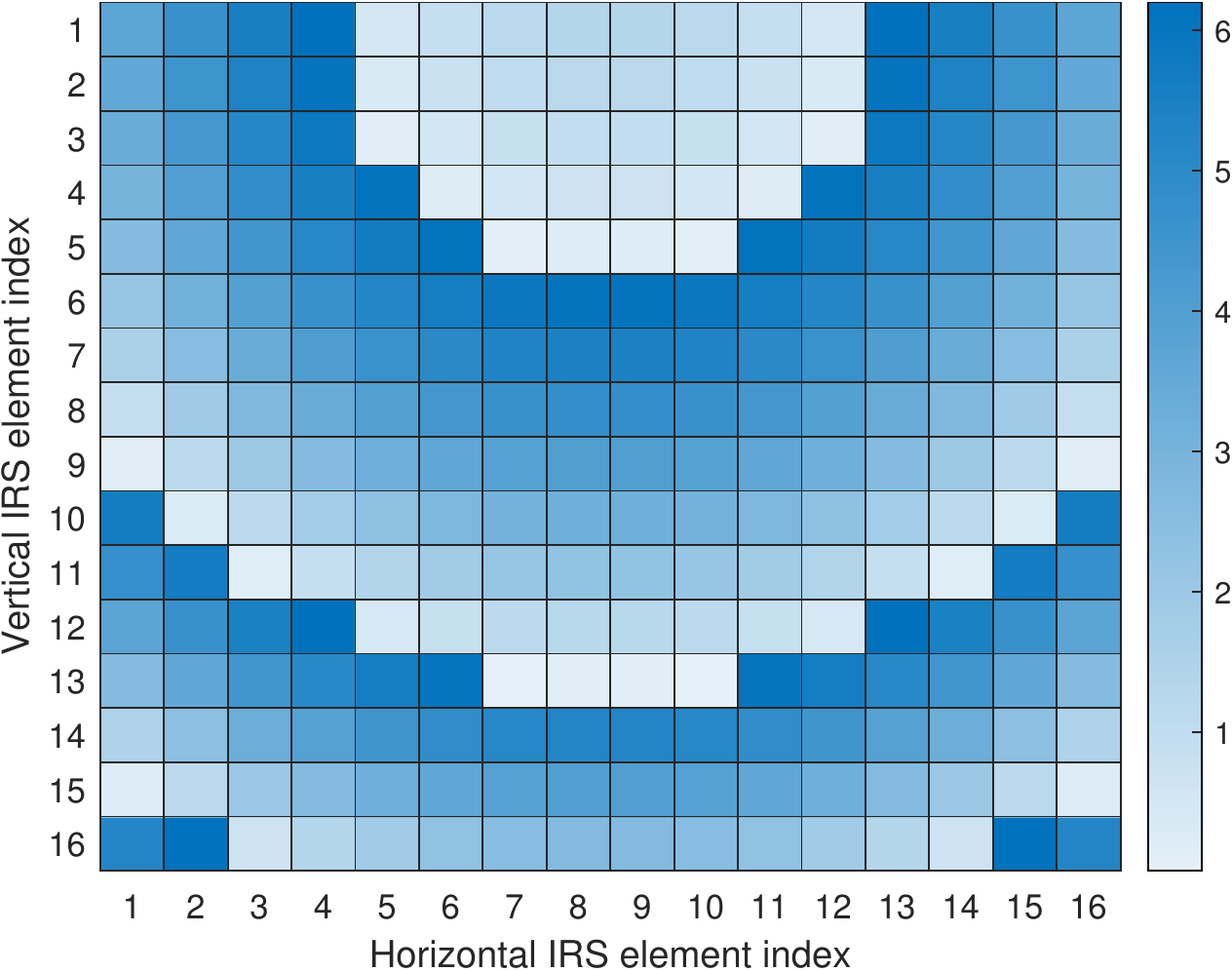}
	\caption{Simulated phase pattern at a $16\times16$ IRS at \SI{5.385}{\giga\hertz} from the geometric path model with an Antenna to IRS distance $d_{\text{IRS}}$ of \SI{1}{\meter}.}
	\label{fig:HeatmapSimulated}
\end{figure}
Figure~\ref{fig:HeatmapSimulated} gives an example of the $2\pi$-periodic phase distribution for a $16\times16$ IRS with a width of ${w=\SI{0.4}{\meter}}$ and a height of ${h=\SI{0.32}{\meter}}$ at a carrier frequency ${f_c=\SI{5.385}{GHz}}$. The transmit and receive antenna is located in the center, at the top of the surface at a distance ${d_\text{IRS}=\SI{1}{\meter}}$.
The resulting pattern of the phase distribution over the surface reflects well the spherical propagation of the wave when it hits the surface.This pattern is a good candidate to achieve the highest possible power of the IRS reflection at the receiving antenna. In order to achieve the highest possible SI
cancellation, the reflections of the individual patches must meet constructively and in phase at the receiving antenna. A good candidate for such a IRS state is thus a projection of the phase distribution, onto a binary state of the surface. The adjustment of the surface thus compensates for the different path lengths and minimizes the SI. 

%% file: GreedyAlgorithm.tex
\section{A Greedy Algorithm for IRS-assisted SI cancellation}\label{sec:GreedyAlgorithm}
\begin{figure}\label{fig:Pap}
	\centering
	\includegraphics[width=0.85\linewidth]{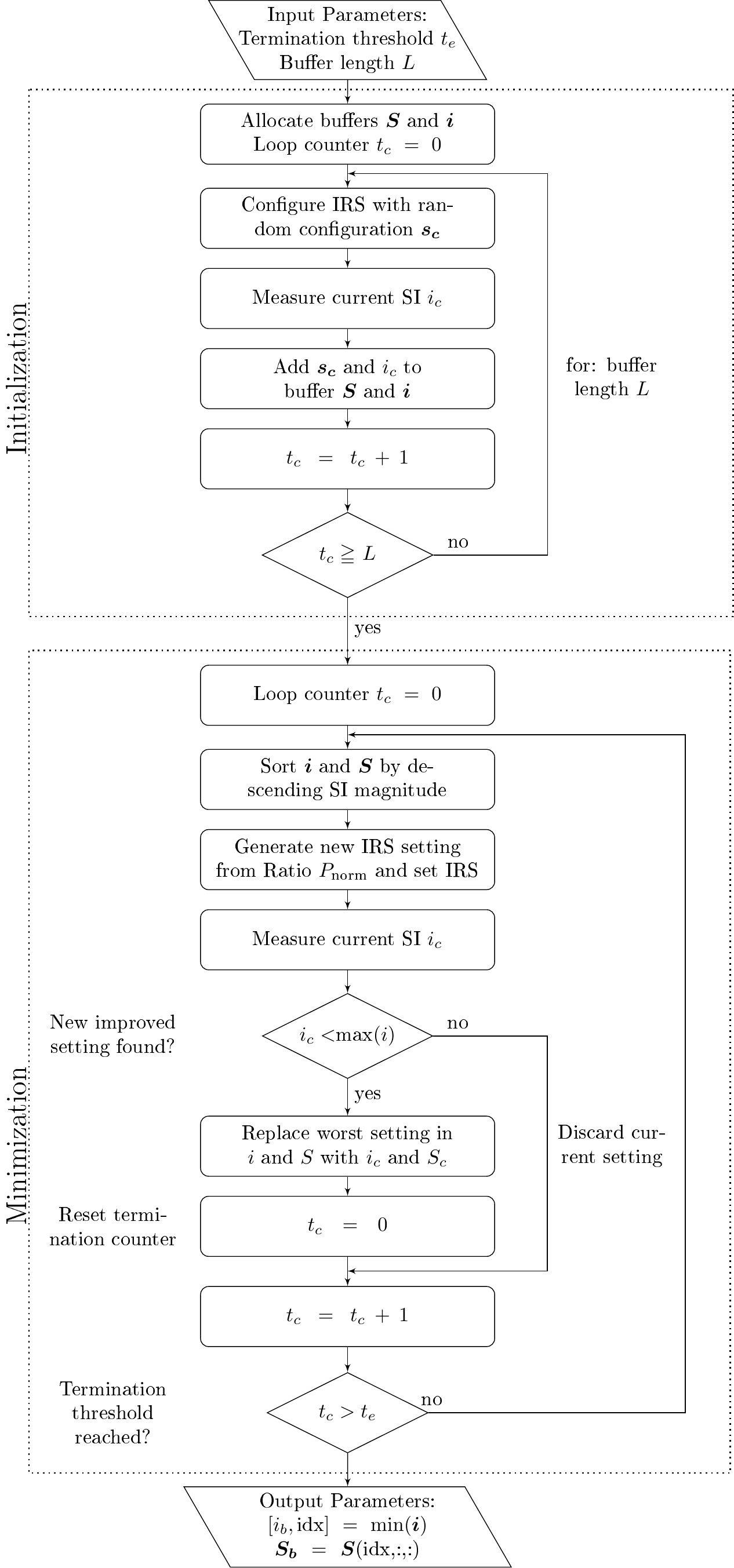}
	\caption{Flowgraph of the proposed greedy algorithm for optimizing the IRS state for SI cancellation.}	
\end{figure}

In this section, we present a greedy algorithm for optimizing SI cancellation. In practical applications, the IRS is part of the infrastructure, or space. A known position of the UE can usually not be assumed, since the manual measurement of it is costly, not very user-friendly and error-prone.
Instead, we propose a greedy algorithm that finds a IRS state within a given number of channel estimations that reduces the SI to a very low level. The proposed algorithm is illustrated in the flow graph in Fig.~\ref{fig:Pap}.
The algorithm is divided into two parts, one is the initialization, which performs some initial random measurements, and the other is the actual minimization phase.
At the beginning, the two buffers, $\mathbold{S}$ and $\mathbold{i}$ are created. The buffer length $L$ is an input parameter of the algorithm and determines the size of the applied buffers. In our experiments, a value of 100 turned out to be reasonable for $L$. Depending on the IRS size, a larger value may be useful here. The binary buffer $\mathbold{S}$ has the dimension $L\times N_x\times N_y$, where $N_x$ is the number of elements of the IRS in the horizontal direction and $N_y$ is the number of elements of the IRS in the vertical direction. In $\mathbold{S}$ $L$ states of the IRS are collected, which are continuously updated as the optimization proceeds.
The buffer $\mathbold{i}$ with dimension $L\times 1$ contains the corresponding measured SI magnitudes to $\mathbold{S}$.

In the initialization phase of the algorithm, $\mathbold{S}$ and $\mathbold{i}$ are initially populated with readings. In a loop, whose number of runs is tracked in the loop counter variable $t_c$, a buffer entry is filled in each run until $L$ runs are reached. For this purpose, a random state is first generated for the surface by a uniformly distributed random function and the surface's current state matrix, referred to as $\mathbold{S_c}$, is set accordingly. Subsequently, the self-interference is quantitatively determined by channel estimation and stored in the buffer $\mathbold{i}$. Different metrics can be used to measure the self-interference, such as the receive signal strength indicator (RSSI), the scaled amplitude of the CSI or an actual received power measurement. Depending on the hardware used, the metric available in the respective case can be chosen. After the completed initialization $\mathbold{S}$ is filled with $L$ IRS states and $\mathbold{i}$ with $L$ associated SI magnitudes.

In the optimization phase, $\mathbold{S}$ and $\mathbold{i}$ are sorted in descending order by SI magnitude in $\mathbold{i}$ in the first step. The sorted matrix $\mathbold{S}$ is then linearly weighted so that the state with the highest self-interference gets the lowest weight and the state with the lowest self-interference gets the highest weight. This results in a ratio $P_\text{norm}$ as
\begin{align}
	P_\text{norm}=\frac{2}{L^2+L}\sum_{l=1}^{L}l\cdot\mathbold{S_l},
\end{align}
with $\mathbold{S_l}$, being a subset of $\mathbold{S}$ with dimension $N_x\times N_y$ representing a single IRS state. The estimate $P_\text{norm}$ is used as a measure of how often a single patch element was active, with the weighting giving a higher fraction due to good IRS states than to less good IRS states.
The ratio obtained is now used as a threshold for a uniformly distributed random function to generate a new IRS state. The state is applied to the surface and the SI is measured for the new state. If the measured SI is smaller than the largest value of the SI in the buffer $\mathbold{i}$, the corresponding state in $\mathbold{S}$ and $\mathbold{i}$ is replaced by the new values. If a new value is found, the loop counter $t_c$, is also reset. This process is performed in an iterative loop until the loop counter $t_c$, which is incremented in each run, exceeds a predefined threshold $t_e$. The optimization thus runs until no new IRS states are found for $t_e$ runs.
As a result, the algorithm returns $\mathbold{S_b}$, which is the IRS state for the lowest SI measured.

%% file: wifiCSI.tex
\section{WiFi CSI}\label{sec:CSI}
\begin{figure}
	\centering
	\includegraphics[width=0.8\linewidth]{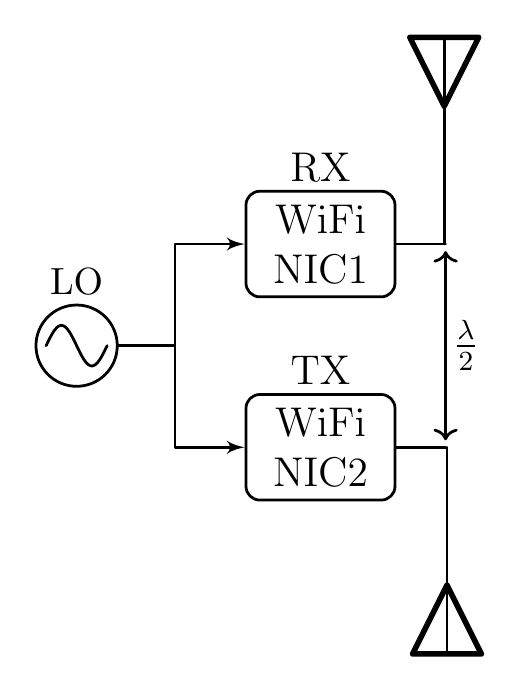}
	\caption{The figure shows the WiFi clocking scheme. Two WiFi NICs are fed a common LO-clock to eliminate CFO and SFO. Both NICs are colocated, one acting as a transmitter, the other as the receiver, respectively. }
	\label{fig:WiFiClockingScheme}
\end{figure}

In our previous work \cite{IRSConf}, we have already demonstrated the basic feasibility of SI cancellation using IRS by means of laboratory hardware, more precisely with a vector network analyzer. However, for a deployment in realistic scenarios, a cheaper and simpler hardware has to be used. In addition to the pure ability to measure the channel, other requirements such as allowed frequency bands and communication must also be ensured.
To address these aspects, we use WiFi, IEEE 802.11n \cite{WiFi-Standard} transceivers in this work. This ubiquitous wireless technology is available at low cost, meets all necessary licensing requirements, and is accepted and widespread among the broad population.
WiFi uses OFDM as the modulation type, so channel estimation is required at the beginning of each WiFi packet for successful data transmission.
At the beginning of each transmitted IEEE 802.11n packet in WiFi, a preamble is first sent containing a short training field (STF) and a periodic sequence, the long training field (LTF). The STF is used for packet start detection and coarse time and frequency synchronization. After the coarse frequency and time synchronization, the LTF is used to obtain a finer estimate of the CFO and sample frequency offset (SFO), and an estimate of the channel. Since the LTF is defined in the IEEE 802.11 WiFi standard, it is known to the receiver. By multiplying the inverse of the known LTF by the received LTF, an estimate of the propagation channel is extracted.
The channel estimate inherent in the WiFi chipset can be extracted with appropriate software tools in certain chipsets. Widely used tools in research are available for Intel NIC 5200 \cite{Intel}, Atheros ATH9k \cite{Atheros} and some Broadcom chipsets \cite{Nexmon}. Meanwhile, some chip vendors are starting to make these CSI available to developers via API interfaces \cite{ESP}.
To capture the phase change of the channel $y_\text{d}(t)$ described in Section~\ref{sec:ChannelModel}, which is caused by breathing, the channel must be measured phase stable between a transmit and a receive antenna. Since WiFi is inherently a half-duplex system, we use two colocated NICs as shown in Fig.~\ref{fig:WiFiClockingScheme}. One network interface card (NIC) acts as a transmitter with one antenna that injects WiFi packets, while the other NIC acts as a receiver at the other antenna and outputs the channel estimate.
The CSI in the receiving device can be written  in the frequency domain for devices with a single antenna as a complex-valued vector $\mathbf{h}$ of dimension $K$, where $K$ is the number of subcarriers used. The elements of $\mathbf{h}$ are given by 	
\begin{align}
h_{k}=|h_{k}|\exp(j \angle h_{k}),
\end{align}	
in which $k$ is the subcarrier index and $\angle h_{k}$ refers to the angle of the complex channel coefficient $h_{k}$.
In a standard WiFi setup, every transceiver derives its LO and sampling clocks from its own crystal oscillators. This results in a number of impairments to the CSI, especially on the phase of $h_{k}$. The estimated phase $\angle\hat{h}_k$ is given by \cite{Atheros,OptimumRecvDesign,OFDMdesign} as
\begin{align}
	\label{equ:3}
	\angle\hat{h}_{k}=\angle h_{k}+(\lambda_{\mathsf{PDD}}+\lambda_{\mathsf{SFO}})\cdot k+\lambda_{\mathsf{CFO}}+\beta+z,
\end{align}	
where $\angle h_{k}$ is the ground truth phase, $\lambda_{\mathsf{PDD}}$ is the phase slope caused by the packet detection delay (PDD), $\lambda_{\mathsf{SFO}}$ is the phase slope caused by the SFO and $\lambda_{\mathsf{CFO}}$ is an additive phase offset caused by the CFO. $\beta$ is a constant system phase offset, and $z$ is the measurement noise. These phase impairments have been extensively studied in various papers \cite{Atheros},\cite{Pi-Splicer},\cite{CalPhaseOffsets}, \cite{CFOtracking}. However, all of these studies looked at individual unsynchronized crystal oscillators in the transceivers.
As has been shown in \cite{IoTj}, there are four major contributors to $\angle \hat{h}_{k}$. The PDD, which results in a linear phase slope with the subcarrier index $k$ and the most severe effect hindering the estimation of a stable phase reading between two WiFi nodes, the CFO. $\lambda_{\mathsf{CFO}}$ is constant for all subcarriers but time varying with every received packet. $\beta$ is a time-invariant phase offset as long as the PLL is locked, i.e., the device is not restarted or the WiFi channel is switched.
To be able to estimate the phase between two WiFi NICs in a colocated antenna assembly, we need to remove $\lambda_{\mathsf{CFO}}$ and $\beta$ to obtain a stable phase reading over several WiFi packets. This is, why we use the architecture proposed in \cite{IoTj}, referred to as wired synchronization for WiFi (WS-WiFi).
In more detail, by feeding two colocated transceivers a common clock, the CFO and SFO are reduced significantly. We modify the WiFi NICs to accept the common wired LO clock, which is applicable in practice since they are colocated by definition in the application, resulting in a new improved $\hat{h}_{k}$ of approx.
\begin{align}
	\angle\hat{h}_{k}=\angle h_{k}+\lambda_{\mathsf{PDD}}\cdot k+z.
\end{align}
In \cite{IoTj}, the LO synchronization method presented was only studied for use in synchronizing multiple WiFi receivers. We use the synchronization procedure instead to synchronize a transmitter and a receiver with each other. Synchronizing the transmitter and receiver allows for phase stable measurement across multiple WiFi packets and is therefore well suited for measuring the $2\pi$-periodic phase between the transmit and receive antennas.

%% file: experimentFloorSetup.tex
\subsection{Breath-Tracking Experiment-Floor setup}\label{sec:ExperimentFloorSetup}

\begin{figure}[!t]
	\centering
	\includegraphics[width=\linewidth]{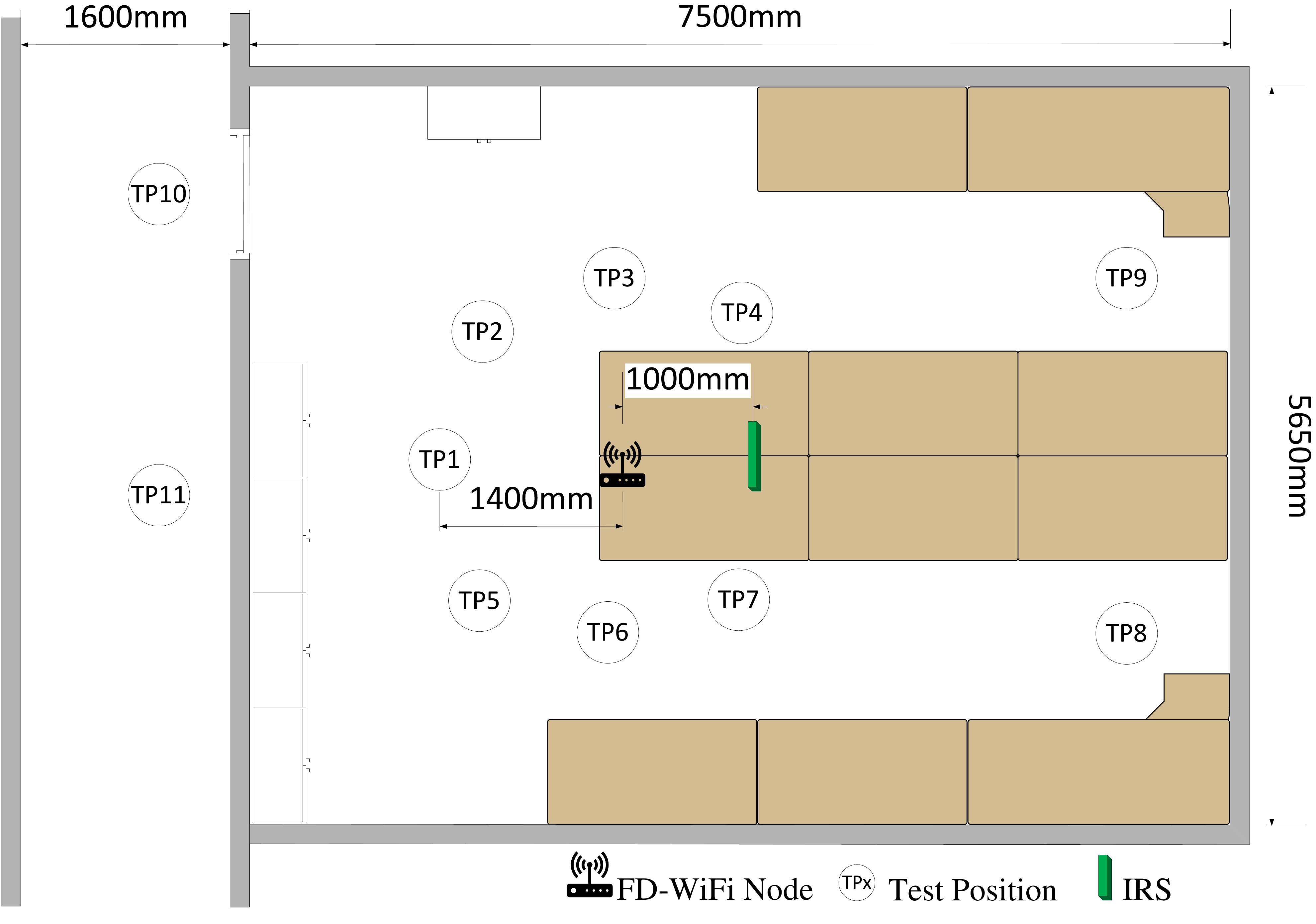}
	\caption{The floorplan of the experiment space shows the dimensions of the room and the relative spacing of test positions and equipment. The test positions refered to during this papers are marked with TP1 to TP11.}
	\label{fig:Floorplan}
\end{figure}

\begin{figure}
	\centering
	\includegraphics[width=0.95\linewidth]{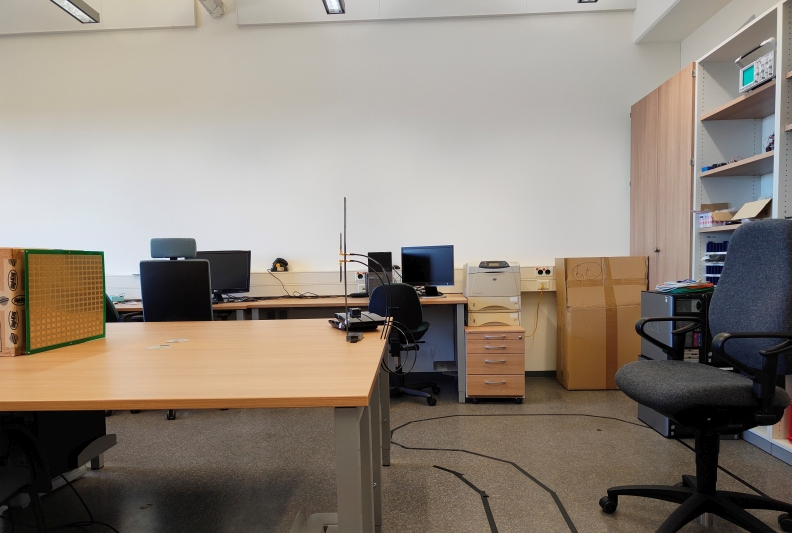}
	\caption{The photograph shows the setup of the breath tracking experiment. On the left the IRS is shown, in a distance of \SI{1}{\meter} to the antennas in the center of the picture. The test person is positioned on the chair at \SI{1.4}{\meter} distance to the antennas to the right at test position 1.}
	\label{fig:Messaufbau}
\end{figure}

The presented experiments were conducted in a typical \SI{42}{\square\meter} office/laboratory environment. The floor plan of the room and the positioning of the equipment are shown in Fig.~\ref{fig:Floorplan}. A photograph of the detailed setup of IRS and antennas is shown in Fig.~\ref{fig:Messaufbau}.
The IRS is set up in the center of the room on a table, with the front of the IRS facing the west direction of the room. At a distance of \SI{1}{\meter} from the surface, the antennas are positioned. The center of the antennas is centered in front of the surface in the top quarter of the IRS.
The room is furnished with typical office furniture, desks, rolling pedestals, and computer workstations. The east front of the room has a full-length window from halfway up, and the south side is adjacent to a hallway. The northern wall and the western wall bordering the corridor are drywall. The remaining walls are load-bearing reinforced concrete walls, as are the floor and ceiling.
A test subject whose breathing is to be detected sits quietly on an office chair with castors and moves to test positions 1 to 11 one after the other for the various measurements. To record the ground truth respiration signal, a NeuLog NUL-236 Chest Strap is used, which records the expansion of the chest via a compression sensor.

%% file: transceiver.tex
\subsection{Transceiver System and Antenna Assembly}
The transceivers we use in our experiments are commodity WiFi transceivers that comply with the IEEE 802.11n standard. More specifically, we use Atheros ATH9k based WiFi routers of the type TP-Link N750 or WDR4300. The routers were patched using the Atheros CSI tool \cite{Atheros} and run OpenWrt, a widely used open source operating system for WiFi routers. The routers we use each offer three antenna ports, but we only use one of each. Thus, a variety of alternative, less expensive WiFi chipsets that support only one antenna can be used. We use two routers, one as a transmitter, the other as a receiver. The transmitter is operated as an injector, i.e., it is not integrated into an access point or client network and can thus inject packets into the wireless medium at a constant rate. The receiver is operated in monitor mode and extracts the CSI of all packets on the set channel that are addressed to its MAC address. The CSI of the received packets is forwarded as UDP packets to a host computer via Ethernet for recording, analysis and coordination. Matching the design frequency of the intelligent surface, WiFi channel 64, with a center frequency of \SI{5.32}{\giga\hertz} at a channel bandwidth of \SI{20}{\mega\hertz} is used.

The channel estimate contains 56 complex values for data subcarriers. The unoccupied guard bands and zero subcarrier are not considered. To ensure stable phase measurements between the transmit and receive antennas, LO synchronization was added to the transceivers. For this purpose, the internal 40 MHz crystal oscillators of the transceivers were removed and replaced by a coaxial connection as described in WS-WiFi~\cite{IoTj}. An external 40 MHz clock is provided by a DG4162 function generator from the manufacturer Rigol and distributed to both routers using coaxial cables. This ensures that no CFO exists between the transmitter and receiver, since both transceivers generate their carrier frequency from the same clock using PLL. The PLL does add phase noise, but this can be removed very well by a low-pass filter over a longer observation period due to the closed loop fashion of the PLL \cite{PLLsurvey}.

In order for the reflection of the IRS to lower the SI by a significant amount, an initial isolation in the analog domain must be realized to approximately compensate the path loss to and from the IRS. For this purpose, e.g., RF circulators can be used, which combine transmitter and receiver on a common antenna. However, these analog RF devices are expensive, difficult to integrate, and require a very well matched antenna, or reconfigurable matching circuit \cite{FDantennaMatching} . The orders of magnitude of isolation that can be achieved with a circulator are in the range of \SI{20}{\decibel} to \SI{30}{\decibel}, but these can only be achieved under ideal conditions with a perfectly matched antenna. We conducted some experiments for this purpose with a DiTom D3C3060-6 circulator with typical Vert 2450 WiFi antennas from Ettus Research, where we were only able to achieve \SI{11}{\decibel} for the analog isolation. 

\begin{figure}
	\centering
	\includegraphics[width=0.85\linewidth]{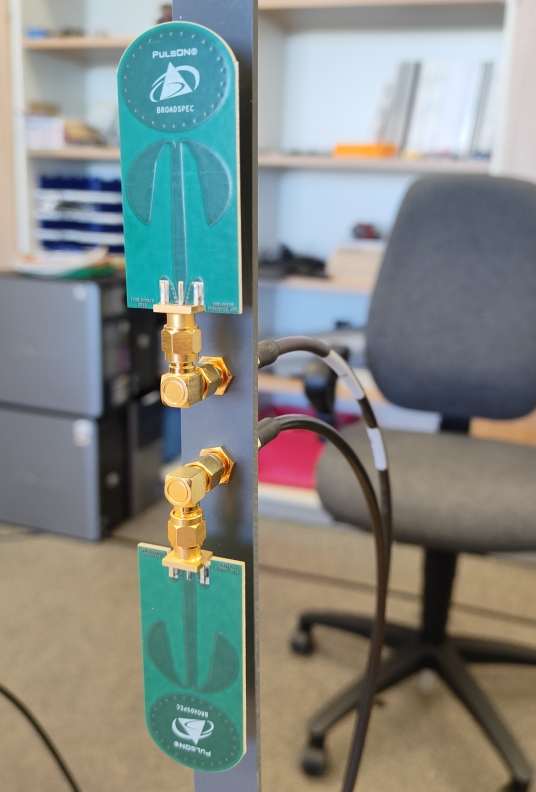}
	\caption{Detail photography of the antenna assembly. The uwb antennas of manufacturer Time Domain Inc. are mounted with $\lambda/2$~distance and are tilted outwards from one another. The transceivers are connected with \SI{1}{\meter} coaxial cables.}
	\label{fig:AntennaCloseUp}
\end{figure}

\begin{figure}[!htb]
	\centering
	\begin{minipage}{.48\linewidth}
		\centering
		\includegraphics[width=\linewidth]{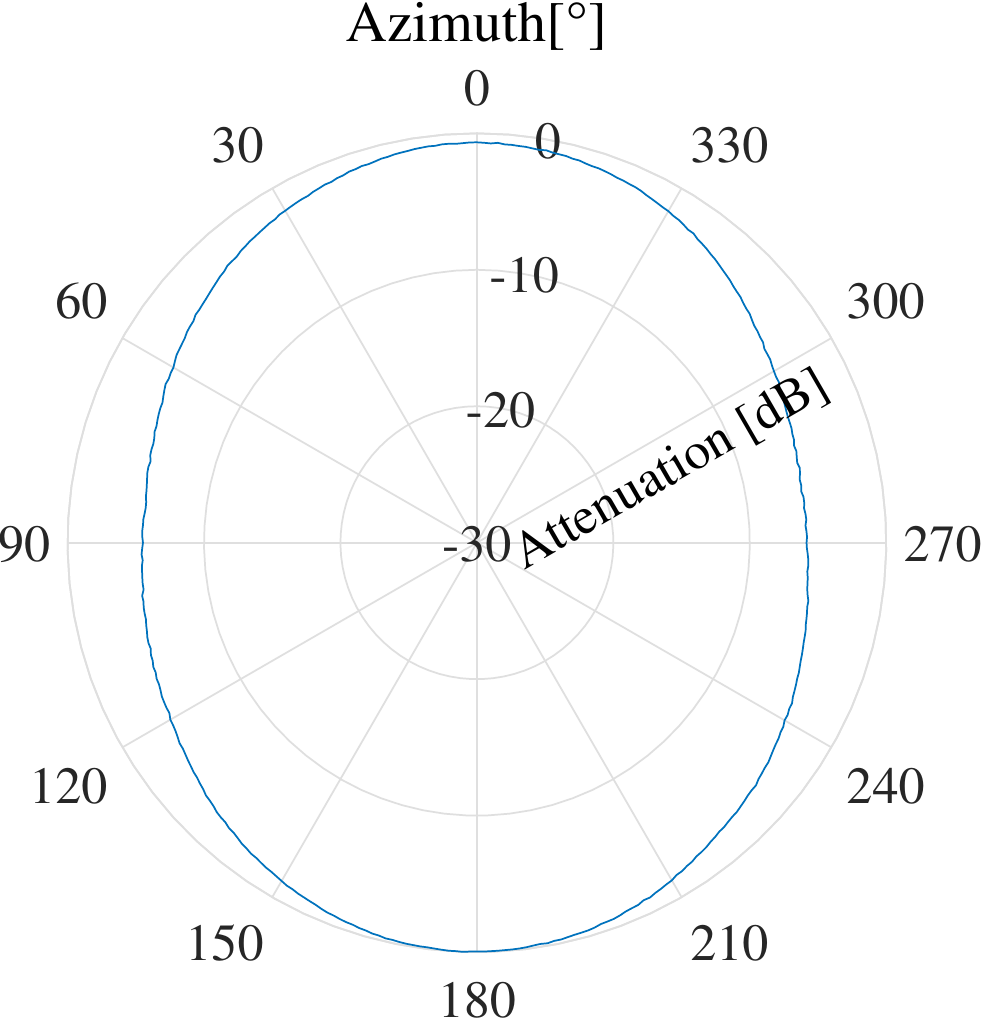}
		\caption{Normalized azimuth antenna characteristic for vertical polarization \SI{5.38}{\giga\hertz}}
		\label{fig:AntennaAzimuth}
	\end{minipage}%
	\hfill
	\begin{minipage}{0.48\linewidth}
		\centering
		\includegraphics[width=\linewidth]{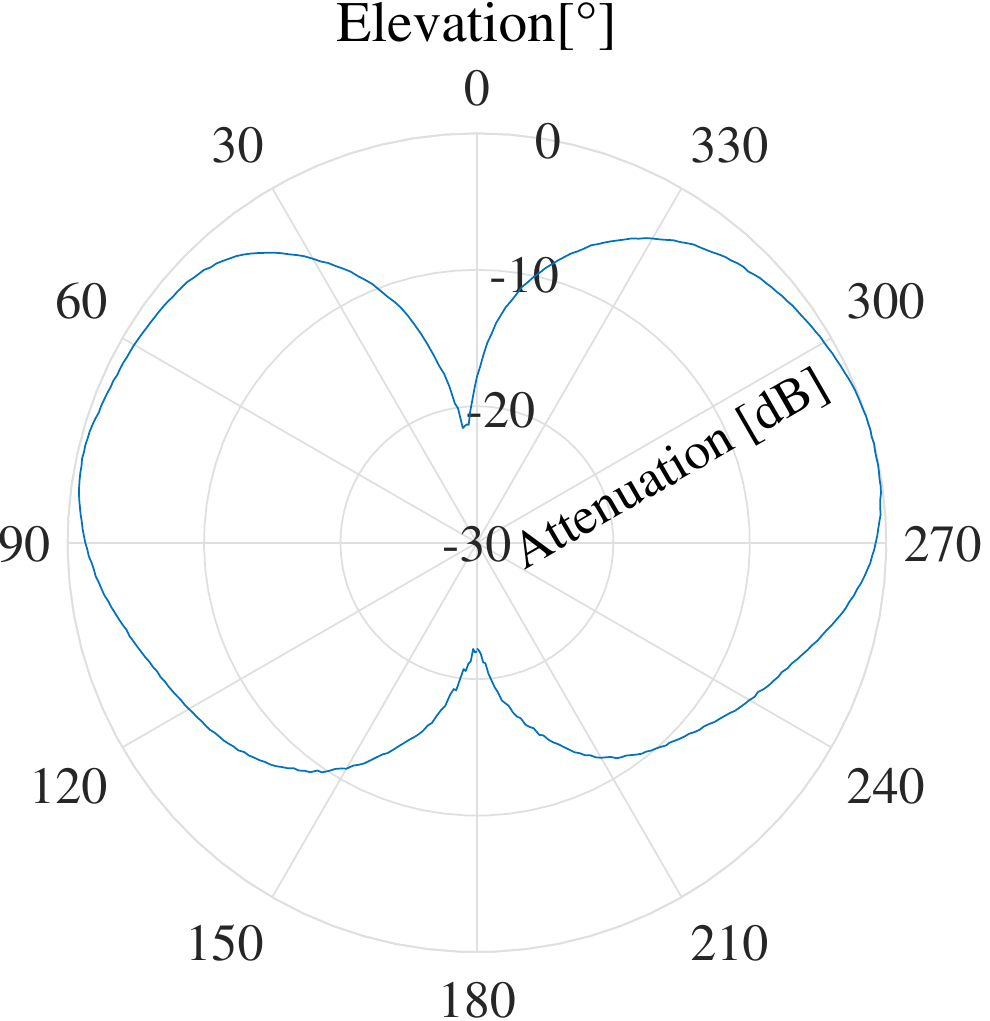}
		\caption{Normalized elevation antenna characteristic for vertical polarization \SI{5.38}{\giga\hertz}}
		\label{fig:AntennaElevation}
	\end{minipage}
\end{figure}

We therefore use an antenna assembly with separate transmit and receive antennas, which we mount at a distance of $\lambda/2$, which is realistic for a UE and easy to implement. A picture of the antenna assembly is shown in Fig.~\ref{fig:AntennaCloseUp}.
The antennas of type Broadspec from Time Domain Inc. are mounted in a polyacetal (POM) plastic bracket using SMA angle adapters and point away from each other. Thus the elevation angles of 180° of the antennas point towards each other.
The actual antenna characteristics were measured in an anechoic chamber for the antennas used. The normalized antenna characteristics are given in Fig.~\ref{fig:AntennaAzimuth} and Fig.~\ref{fig:AntennaElevation}.

The antenna attenuation at elevation angle 180° can be read as \SI{-22}{\decibel}. Since both antennas used have this characteristic, an analog isolation between transmitter and receiver of \SI{44}{\decibel} could be achieved if the antennas are ideally aligned with each other. In our experiments, we could typically achieve values of \SI{40}{\decibel} in an indoor environment, which is sufficient for the intended use with the available IRS. The analog isolation, may now seem quite high for embedding in a UE, such as a smartphone, however this value may be relaxed in the future with the availability of larger IRSs.

In the arrangement of the antennas shown, the test subject to be detected is thus located in the azimuth plane of the antennas. If we examine the antenna pattern in Fig.~\ref{fig:AntennaAzimuth}, we observe that the antenna pattern is very good in all directions and almost has an omnidirectional characteristic. The maximum attenuation is \SI{5}{\decibel} at the side-view of the antennas. This ensures that there are no blind spots and that 360° coverage around the antenna assembly is possible.

%% file: irs.tex
\subsection{The IRS}\label{sec:irs}
\begin{figure}
	\centering
	\includegraphics[width=\linewidth]{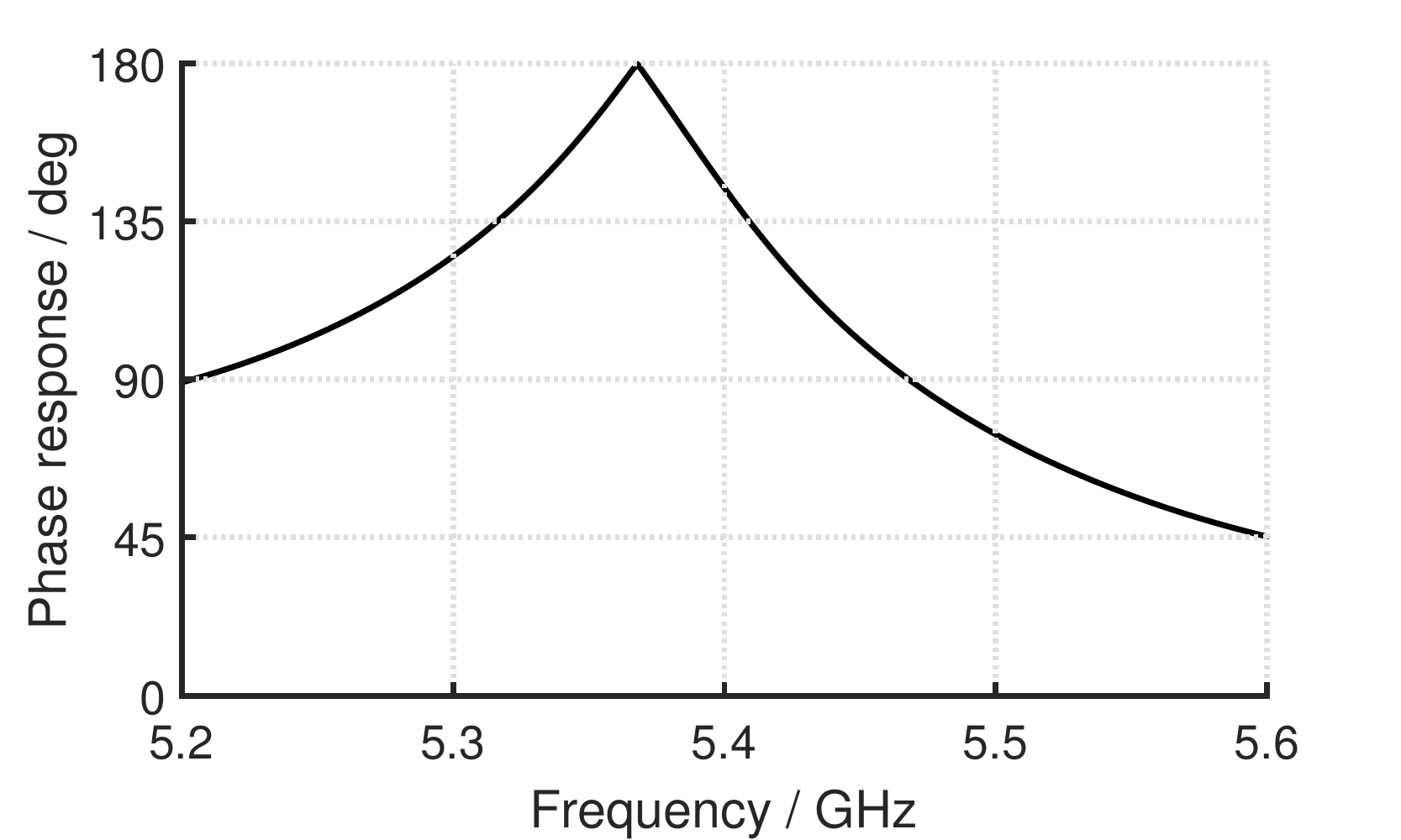}
	\caption{Measured phase response of the IRS.}
	\label{fig:IRSPhaseResponse}
\end{figure}

We use an array of $16 \times 16$ identically structured unit cells embedded on a printed circuit board (PCB) as our IRS prototype. Each unit cell offers a binary switchable resonance frequency, which results in two switchable reflection coefficients with different phase. The unit cell consists of a vertically polarized rectangular patch reflector on the top side of the PCB and a contiguous ground plane on the bottom side. The length of the patch is in the order of half a wavelength and therefore defines the resonance frequency of the patch reflector in the unloaded state. A copper-plated through hole connection (via) connects the patch on the top side to the anode of a PIN diode located on the bottom side of the IRS. The cathode of the PIN diode is connected to the ground plane. An eccentric position of the via affects the electrical length of the patch when the PIN diode is biased. Thus, the resonance frequency of the patch reflector can be altered by shorting the via to ground. This causes a binary-switchable resonance frequency of the unit cell. The resulting phase response of the IRS is shown in Fig.~\ref{fig:IRSPhaseResponse}, which is the phase difference of the surface’s reflection coefficient between the IRS configured as all-ON and all-OFF states. It was measured with the reflected wave vector and the incident wave vector perpendicular to the surface.

\begin{figure}
	\centering
	\includegraphics[width=\linewidth]{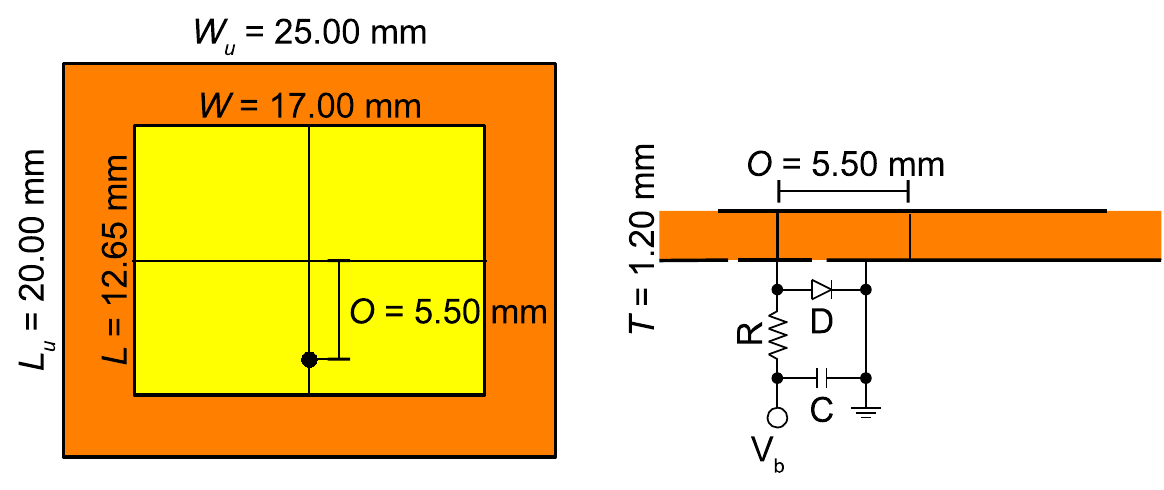}
	\caption{Physical dimensions, and driver circuit of an IRS unit cell.}
	\label{fig:UnitCell}
\end{figure}

Fig.~\ref{fig:UnitCell} shows the physical dimensions and driver circuitry of the unit cell. Standard FR4 material with a dielectric constant of $\epsilon_r \approx 4.5$ and a dissipation factor of $\tan\delta \approx 0.02$ is used as substrate. A PIN diode from manufacturer Skyworks of type SMP1320-079LF is used. The forward current through the PIN diode of $I_f = \SI{1}{\milli\ampere}$ is set by a biasing resistor of value $R = \SI{4.7}{\kilo\ohm}$. A capacitor of value $C = \SI{100}{\pico\farad}$ is used for additional decoupling of unwanted AC signals. In the ON state or the OFF state, a biasing voltage of $V_b = \SI{5}{\volt}$ or $V_b = \SI{0}{\volt}$ is applied, respectively. This unit cell is an evolution of our prior work from \cite{Heinrichs}.

The IRS can be controlled as follows: ASCII-coded text commands, containing the surface state to be configured, are send to a microcontroller located on the back side of the IRS via a universal asynchronous receive-transmit (UART) interface. The microcontroller then generates a binary sequence, which is shifted into 32 daisy-chained 8-bit shift registers. Each unit cell is connected to one of the 256 outputs of the shift registers, which supply the biasing voltages to the array elements. We use the following components: 74HC565D shift registers from NXP Semiconductors, an STM32F446RET6 microcontroller from STMicroelectronics running at 180 MHz, and an MCP2221A from Microchip Technology as a USB-to-UART bridge.

With the selected PIN diode and the implemented biasing circuit, we measured a minimum switching time of approx. 4 µs to switch a unit cell from the ON-state to the OFF-state. This is the time required for the PIN diode to become non-conductive for RF signals. Thus, a maximum of approx. \SI{250}{\kilo\relax} IRS states can be configured in one second. With the shift registers used, one can attain a maximum of approx. \SI{390}{\kilo\relax} configurations per second. Thus, the PIN diodes limit the configuration rate achievable, which affects the time required to find a suitable IRS state for cancellation of self-interference. To provide a preview of our future work, we measured the configuration rate of a unit cell with a modified biasing network. By omitting the decoupling capacitor C and adding a parallel resonance circuit across the biasing resistor $R$, which decouples the control circuit from the unit cell, a switching time of approx. \SI{40}{\nano\second} can be achieved. This leads to configuration rates of \SI{25}{\mega\relax} configurations per second, if the proposed biasing network is modified and the digital control circuitry is improved for fast switching. In this case, a control interface offering much higher data rates than UART has to be used.

%% file: phaseStability.tex
\subsection{Phase Stability of the setup}\label{sec:PhaseStability}
To verify that the LO synchronization approach presented in \cite{IoTj} also works as expected for our case of synchronized transmitters and receivers, we first check the phase stability of the system used. For this purpose, the transmit and receive antennas of the antenna assembly are detached and replaced by a connection with a coaxial cable and a \SI{30}{\decibel} attenuator. This ensures constant phase over a long period of time, which we examine below.

\begin{figure}
	\centering
	\includegraphics[width=.9\linewidth]{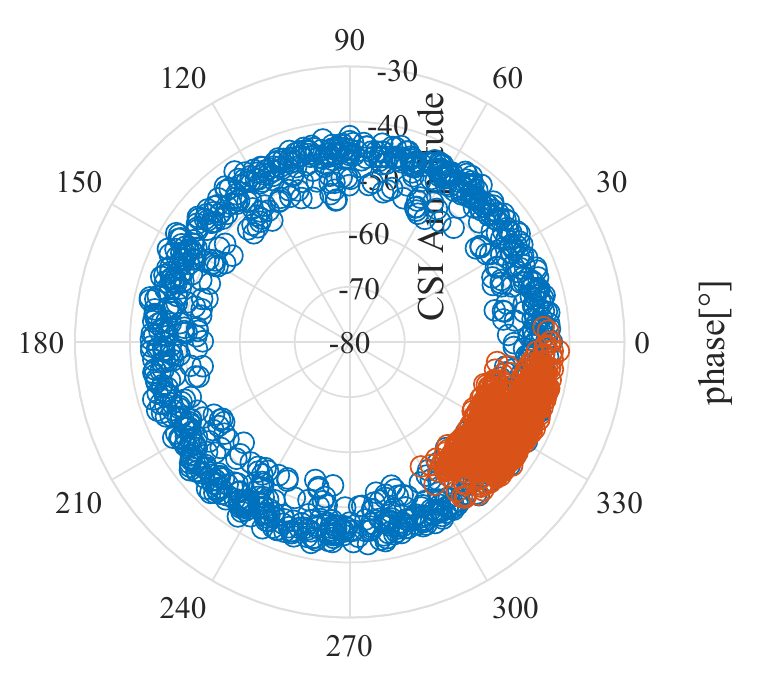}
	\caption{Polarplot of the 25th subcarrier's CSI phase over a time of \SI{10}{\second} at \SI{400}{\hertz} packet rate}
	\label{fig:PhaseScatter}
\end{figure}

The transmitter injects WiFi packets at a rate of \SI{400}{\hertz}, which are received by the receiver and the CSI is evaluated. We consider subcarrier 25 for the phase stability study. The phase of the subcarrier is recorded over a period of \SI{10}{\second} and plotted as a polar plot in Fig.~\ref{fig:PhaseScatter}.
\begin{figure}
	\centering
	\includegraphics[width=.9\linewidth]{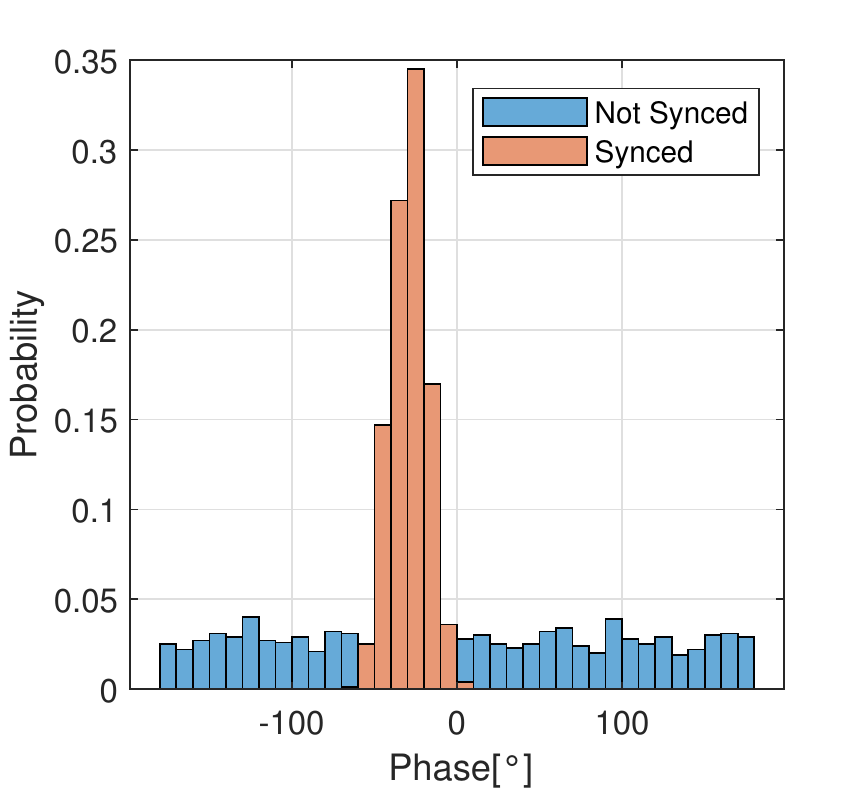}
	\caption{Histogram of the phase measurements}.
	\label{fig:PhaseHistogram}
\end{figure}
For this initial test at room temperature, we compare two measurements of 10s each. The measurement shown in blue in the polar plot was performed without LO synchronization. It can be seen that the phase of the subcarrier under consideration does not provide stable phase information across several WiFi packets. With LO synchronization applied, we obtain the measurement shown in red. It can be seen that with the applied synchronization a stable phase measurement is obtained with a maximum scatter of about 60°. The scattering, which is significantly added by the phase noise of the PLL \cite{PLLsurvey}, is only a short-term effect, which is averaged out over a longer measurement period, since a long-term drift is not possible due to the closed-loop fashion of a PLL.

 In Fig.~\ref{fig:PhaseHistogram} the distribution of the phase for the same experiment is depicted. It can be seen, that for the red curve, with synchronization the distribution centers at a stable value with small bandwidth. The phase noise can therefore be very well compensated for slow measurands, such as respiration with a typical frequency of below \SI{1}{\hertz} can be filtered very well by a moving average filter or a low pass filter. Since PLLs have a significant temperature dependence, this can introduce limitations of the system for certain applications. We perform an analysis of the phase stability of the 25th subcarrier over an extended temperature range from \SI{-20}{\degreeCelsius} to \SI{40}{\degreeCelsius}. For this purpose, the transmitter's SMA port is still connected to the receiver's SMA port via a coaxial cable to ensure a stable phase in the analog domain. The test is performed in two configurations. Once a router is placed in a thermal chamber of type VT4002 from manufacturer Vötsch Industrietechnik and a router is placed at constant room temperature to obtain a maximum temperature difference. In the second test, both transceivers are placed in the thermal chamber so that they are kept at the same temperature and only experience an ambient temperature difference, but there is no significant temperature difference between the two devices. The ambient temperature is recorded by sensors in the chamber and in the room, respectively. In addition, the temperature of the WiFi chipsets is recorded by a DS18B20 temperature sensor from the company Dallas Semiconductor, which is glued directly onto the chipset.
\begin{figure}
	\includegraphics[width=\linewidth]{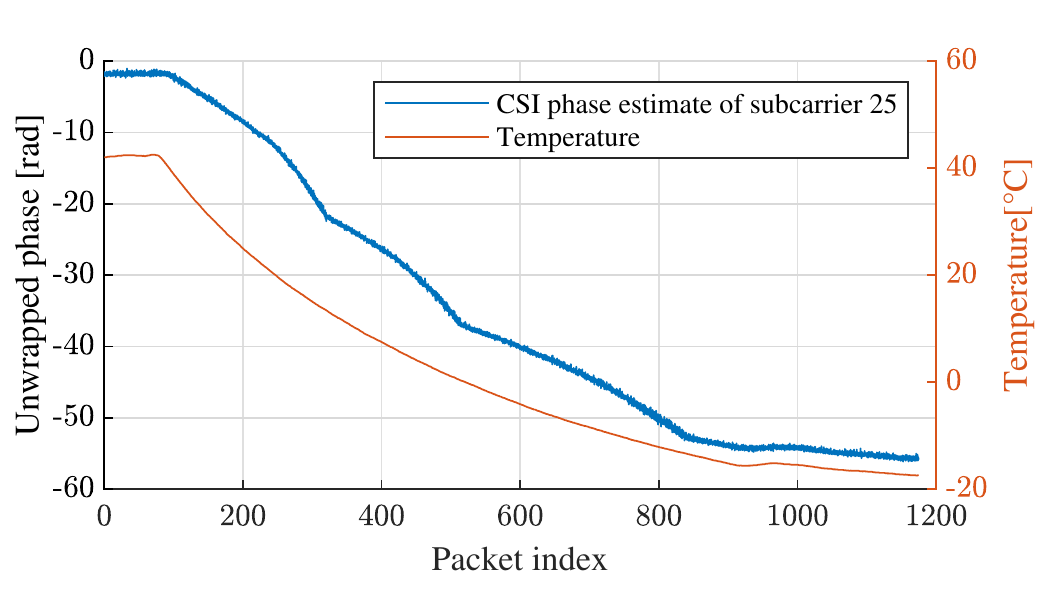}
	\caption{Phase change of subcarrier 25 over a full temperature cycle from \SI{40}{\degreeCelsius} to \SI{-20}{\degreeCelsius} with one transceiver at constant \SI{22}{\degreeCelsius} room temperature and the other in the thermal chamber.}
	\label{fig:TempDependency}
\end{figure}

In Fig.~\ref{fig:TempDependency}, the measured phase of the 25th subcarrier is plotted versus temperature. A complete temperature cycle from \SI{40}{\degreeCelsius} to \SI{-20}{\degreeCelsius} ambient temperature was performed over a time of \SI{20}{\minute} at a packet rate of \SI{1}{\hertz} to ensure sufficient acclimation to the continuous slow temperature drop in the chamber. Temperature alignment was ensured by comparing the on-chip temperature sensors and the ambient temperature sensors. These had a constant relation to each other throughout the measurements with a maximum deviation of \SI{2.3}{\degreeCelsius}.
In the plot, the phase change over the chamber ambient temperature is shown in blue, where the transmitting router was operated outside the chamber, at a constant ambient temperature of \SI{22}{\degreeCelsius}. Shown in red is the temperature throughout the experiment.
It can be seen that there is a temperature dependence of the measured phase on temperature. 

For the case of breath detection in indoor environments considered here, a rather slow ambient temperature change in a moderate temperature range can be assumed. Furthermore, since the two transceivers are always placed spatially close to each other by the colocated antennas, a very small differential temperature can generally be assumed. Considering the slow change of temperature over several seconds to minutes, the change of respiration is much faster. Since only a relative observation of the phase change over time is relevant for the detection of respiration, the influence by a temperature difference change can be neglected. Furthermore, if the observation period is very long, the slow increase or decrease of the phase difference due to the temperature can be removed by a simple high-pass filter with a cut-off frequency below the respiration frequency.

%% file: algoConvergence.tex
\subsection{Reduction of SI and convergence of the greedy algorithm}
To ensure the effective operation of SI cancellation by the IRS and the convergence to good SI values with the presented greedy algorithm and Atheros 9k based WiFi transceivers, we analyze the measure of SI using the CSI.
The greedy algorithm proposed in section~\ref{sec:GreedyAlgorithm} can use different metrics as a measure of SI. In addition to a relatively coarse receive level measurement using the RSSI, WiFi CSI allows for a subcarrier-based analysis. For this, we use the amplitude of the CSI extracted using the Atheros CSI tool \cite{Atheros}. Since the amplitude is scaled to 10 bits and does not take the AGC gain into account, it is scaled with a normalization using the RSSI and thus a more accurate receive level estimate is achieved than with the pure received signal strength indicator (RSSI). The scaled amplitude of the CSI is now averaged over all 56 subcarriers to obtain the broadest possible SI suppression in the WiFi band and used as the SI estimate for the greedy algorithm.

\begin{figure}[!t]
  	\includegraphics[width=\linewidth]{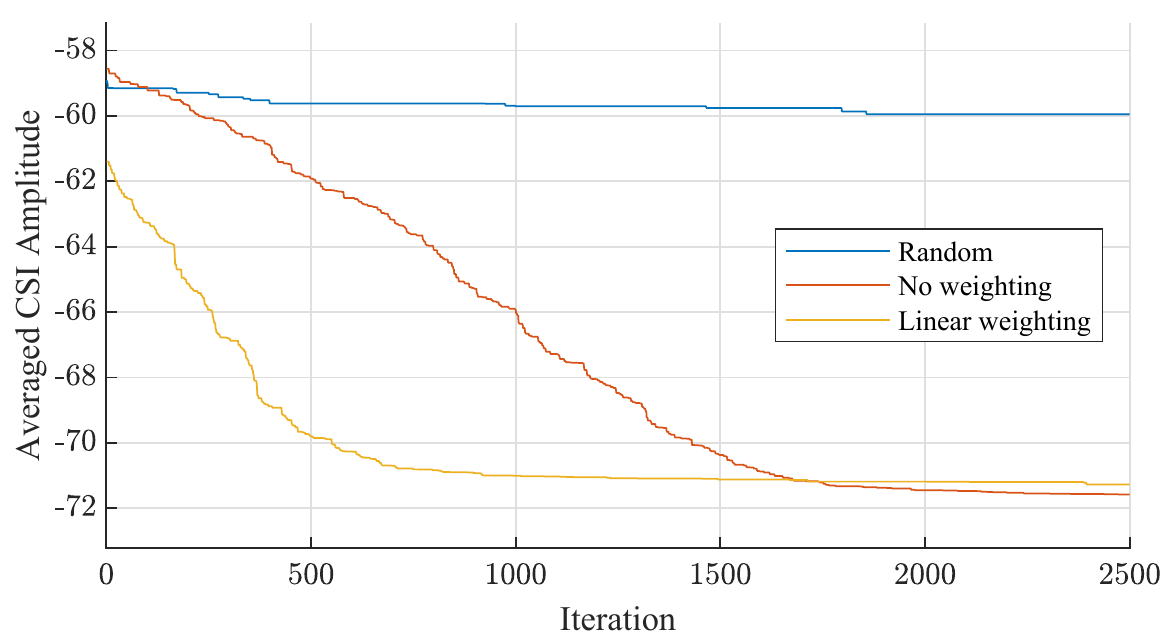}
 	\caption{Cummulative minimum convergence plot averaged from 10 runs. The depicted amplitude is derived from the CSI amplitude, which is averaged from 56 subcarriers in the carrier band.}
 	\label{fig:ConvergencePlot}
 \end{figure}

For verification, we run the algorithm 10 times in the setup described in section~\ref{sec:ExperimentFloorSetup} and track the cummulative minmum during the optimization iterations. Fig.~\ref{fig:ConvergencePlot} shows the convergence averaged over the 10 runs. The diagram shows the convergence~(yellow line) of the algorithm presented in section \ref{sec:GreedyAlgorithm}. With the applied linear weighting, the algorithm reduces the averaged amplitude as a measure of the SI to -~70 after the initialization phase within 500 iterations. If the linear weighting is neglected within the greedy algorithm and all collected values in the buffer $\mathbold{S}$ are considered with the same weighting, the value of -~70 is also reached after 1600 steps, but 1100 more iterations are required than with the linear weighting.
As a baseline, an additional comparison is drawn with a random generation of IRS states as an optimization method. The 'random' case shown in blue shows a barely noticeable improvement in SI over the 2500 steps considered.
A single iteration of the greedy algorithm's optimization loop requires \SI{0.1}{\second} for 2500 optimization iterations hence \SI{250}{\second} are required, which is mainly limited by the execution times of the different interfaces and the maximum update rate of the IRS. With an optimized control of the surface this value can be reduced substantially. The maximum update rate of the IRS limited by the pin diodes specified in section~\ref{sec:irs} is significantly higher than the packet rate achievable with WiFi. 

\begin{figure}[!t]
	\includegraphics[width=\linewidth]{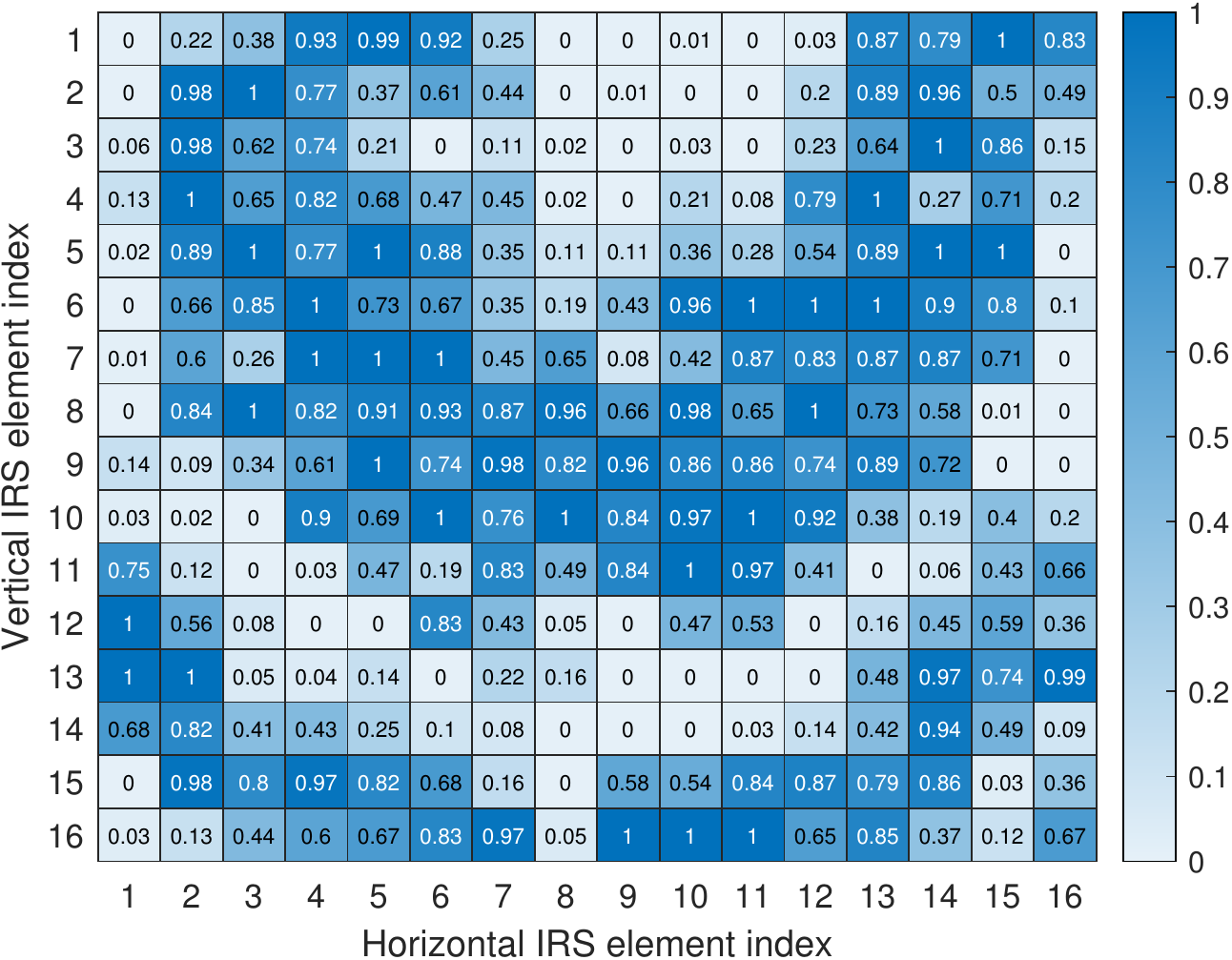}
	\caption{Converged IRS surface state, averaged and normalized from buffer~$\mathbold{S}$, used in the greedy algorithm.}
	\label{fig:HeatmapConverged}
\end{figure}

 Note, since we are working with non-calibrated commodity instruments, the convergence, or SI, cannot be given in absolute power levels, but only as a relative estimate. For a quantitative absolute investigation of the SI, please refer to our previous paper \cite{IRSConf}. 
 After successful execution and convergence of the greedy algorithm, the IRS states buffer $\mathbold{S}$ can be analyzed for a better understanding of the found IRS state. The buffer $\mathbold{S}$ contains the $L$ best states found during all runs at the end of the greedy algorithm. Averaging $\mathbold{S}$ over the $L$ runs and normalizing it to 1 results in the pattern shown in Fig.~\ref{fig:HeatmapConverged} in the form of a heatmap. Here, the values shown represent a ratio of how many times an element was activated in the $L$ states. It is easy to see a strong similarity with the phase distribution in Fig.~\ref{fig:HeatmapSimulated}. This commonality makes sense since the pattern balances the geometric path lengths as much as possible and results in a constructive superposition with the correct phasing at the receiving antenna. 
 The obtained values for SI cancellation show a significant improvement over the non-optimized case, and the SI-optimized transceiver system thus obtained will be studied in the next subsection for respiration tracking application.
 
%% file: breathTracking.tex
\subsection{Experimental Results of IRS-assisted Breath Tracking}
For application to the acquisition of respiration from a test person, the system described previously was evaluated. To capture respiration, the phase of the subcarrier 25 of the received CSI data is examined and presented below. In addition, for comparison with the actual respiration pattern, a NeuLog NUL-236 chest strap sensor is used. The chest strap measures respiration using a rubber balloon that is compressed as the chest expands and is evaluated by a pressure sensor. However, since the measured pressure values of the chest belt strongly depend on the pre-pressure on the chest belt and the tightness of the belt, the amplitude is normalized to 1 over the measurement period in the following. Since the envelope of the respiration signal is essentially relevant for the recording of the respiration, no information is lost as a result, but a more consistent, more comparable ground truth reading is obtained.

\begin{figure}
	\centering
	\includegraphics[width=\linewidth]{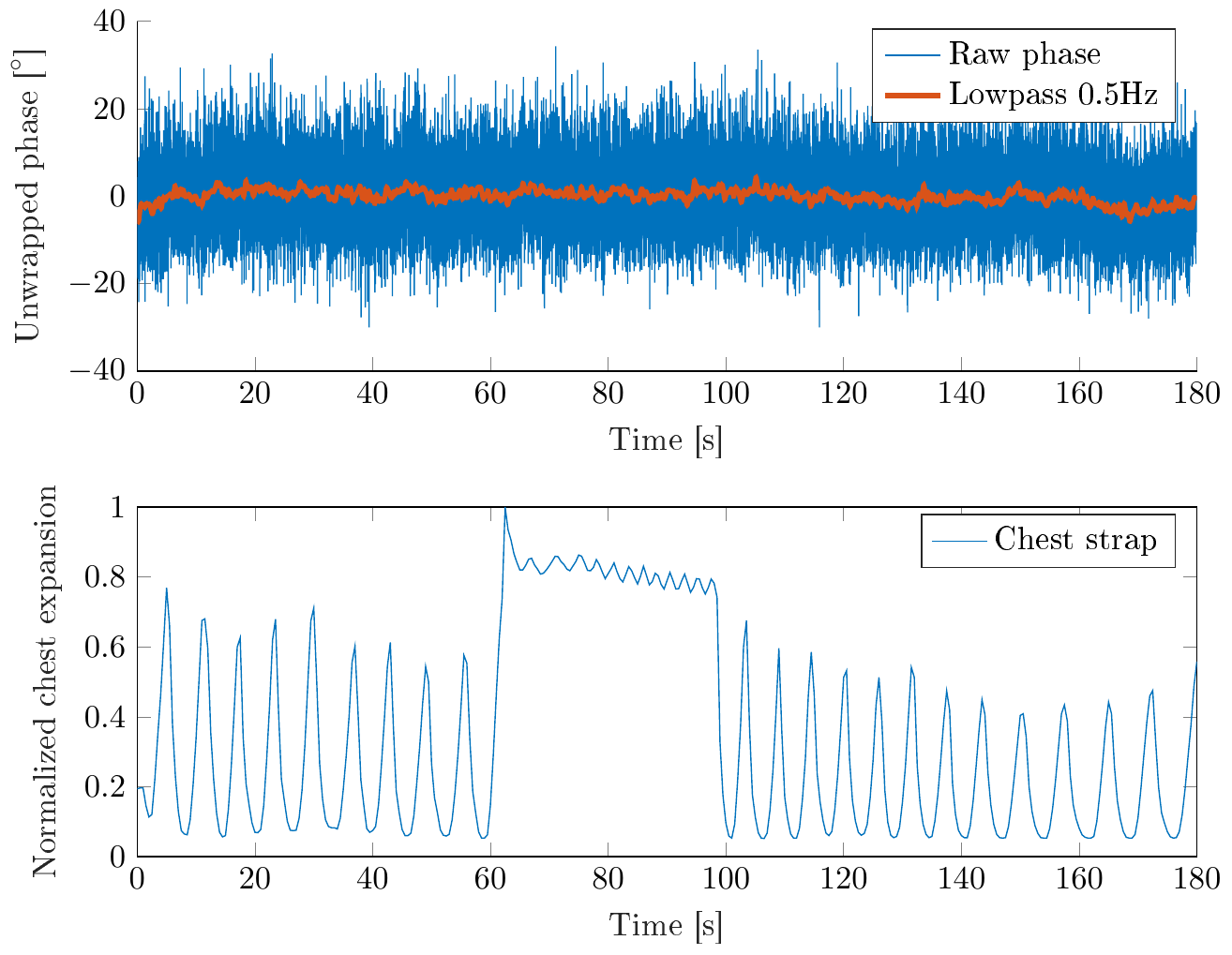}
	\caption{Breathing signal for test postion 1 from the chest belt in the lower plot and  from the proposed setup without IRS optimization in the upper plot.}
	\label{fig:BreathingNonOptimizedTP1}
\end{figure}

As an initial measurement, a measurement without optimization of the IRS was performed. For this purpose, the surface was set to the 'all elements off' state and a measurement was recorded over the period of \SI{180}{\second} at test position 1. During the measurement period, the subject sat quietly on an office chair and breathed regularly at the beginning, held its breath after a few breaths, and then continued to breathe regularly. The results are shown in Fig.~\ref{fig:BreathingNonOptimizedTP1}. The bottom plot of the diagram shows the ground truth breath signal measured by the chest strap. The upper plot shows the raw measurement of the phase on subcarrier 25 in blue and the phase filtered using a low-pass filter at \SI{0.5}{\hertz} cutoff frequency is shown in red. The packet rate for the phase measurement is \SI{400}{\hertz}.

\begin{figure}
	\centering
	\includegraphics[width=\linewidth]{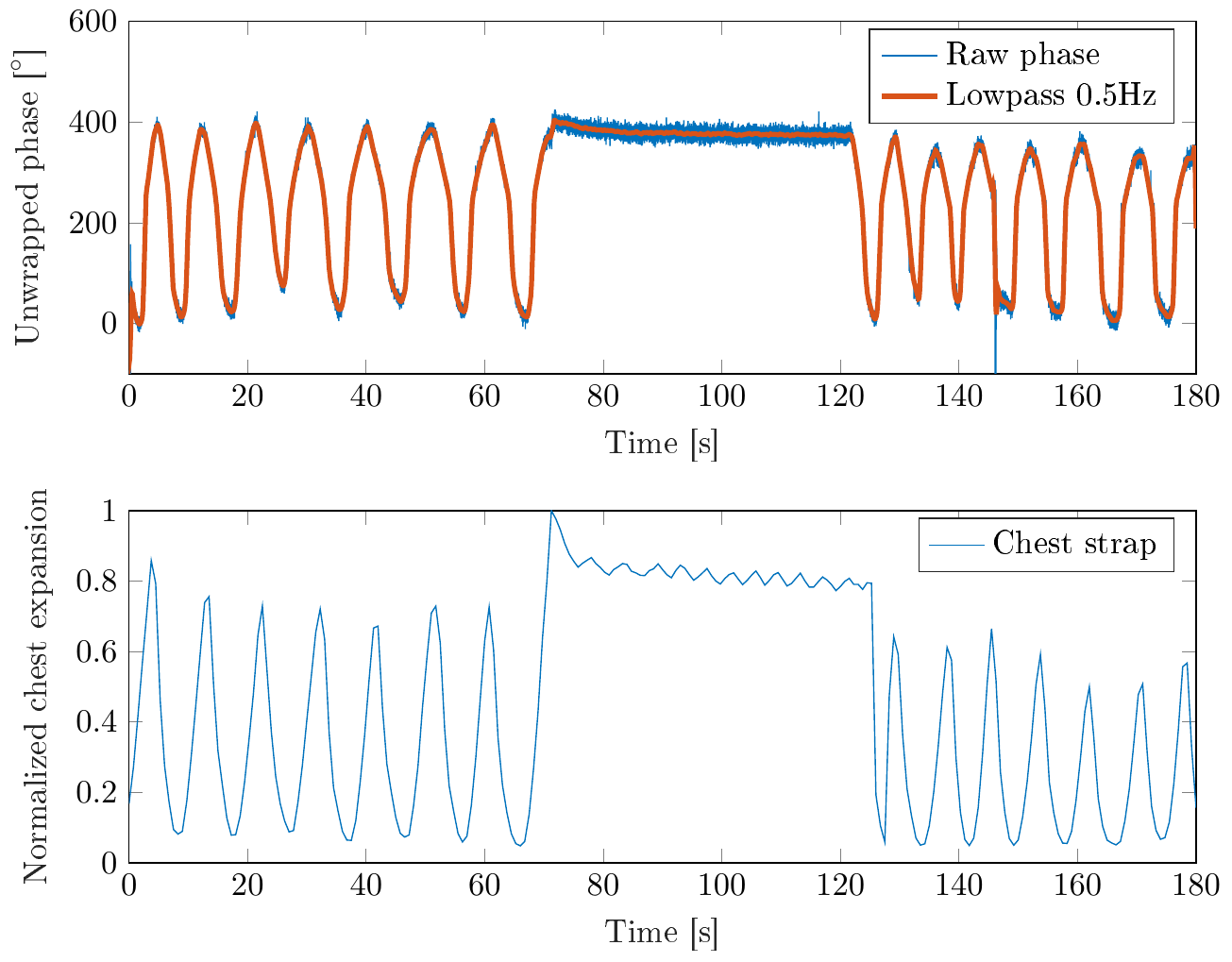}
	\caption{Breathing signal for test postion 1 from the chest belt in the lower plot and  from the proposed setup with IRS optimization in the upper plot.}
	\label{fig:BreathingTP1}
\end{figure}

The raw phase signal shows noise in the range -20° to 20°, which, as also shown in section~\ref{sec:PhaseStability}, is caused by the phase jitter of the PLL of the measurement system. However, long-term stability is given, so by applying the \SI{0.5}{\hertz} low-pass filter to limit it to the range of a typical respiratory signal, the short-time phase jitter is also removed. The resulting signal contains little phase change and cannot be used to detect respiration.
We now apply the optimization using the greedy algorithm. The algorithm is executed once at the beginning and the IRS state found is retained for the measurements. The same breathing pattern of regular breathing, holding breath and regular breathing again was measured. The results are shown in Fig.~\ref{fig:BreathingTP1}. The lower plot again shows the actual breathing pattern using the chest strap. The top plot again shows the raw phase signal and the low-pass filtered phase of the subcarrier 25.

The phase plot shows a much larger change, of up to 400° and traces the respiration signal very clearly. It is readily apparent that by means of IRS-assisted reduction of the SI, detection of respiration is made possible. It also becomes possible to detect not only the frequency of respiration, but also the pattern of respiration. The plot demonstrates for the first time respiration detection with commodity WiFi devices and a single colocated transceiver site in an indoor environment.

The phase pattern shown in Fig.~\ref{fig:BreathingNonOptimizedTP1} and Fig.~\ref{fig:BreathingTP1} is exemplary for test position 1, the remaining test positions 2 to 7 in \SI{1,40}{\meter} radius of the antenna setup show a comparable pattern and are not broken down in detail. We can confirm that there is no significant directional dependence of the respiration detection and that 360° detection is possible.
Test positions 8 and 9 show a comparable pattern to test position 1, however the absolute phase change has a smaller value of 110° maximum. Breathing can still be detected without any problems. The test positions 10 and 11 in the corridor no longer provide meaningful results. Even though a phase change could be detected at TP10, it is no longer visible at TP11 due to the high attenuation of the drywall and cabinets.
We have demonstrated in our experiments that respiration detection of a single person sitting quietly at room level range is possible with commodity WiFi hardware and a supporting IRS with only a single co-located antenna site.